\documentclass[arxiv]{acmart}

\usepackage{listings}
\usepackage{todonotes}
\usepackage{subcaption}
\usepackage{algorithm}%% added by yingjie
\usepackage[noend]{algpseudocode}
\usepackage[normalem]{ulem}
\usepackage{chemarrow}
\usepackage{xltabular} %% added by yingjie
\usepackage{xspace} % added by di for xspace in text command
\usepackage{hyperref} % added by di
\newtheorem{theorem}{Theorem}

\newtheorem{definition}{Definition}

\newcommand{\Swap}{\ensuremath{\mathit{Swap}}}
\newcommand{\Xcoin}{\ensuremath{\mathit{Xcoin}}}

\newcommand{\Ycoin}{\ensuremath{\mathit{Ycoin}}}

\newcommand{\current}{\ensuremath{\mathit{current}}}
\newcommand{\Escrow}{\ensuremath{\mathit{Escrow}}}
\newcommand{\Redeem}{\ensuremath{\mathit{Redeem}}}
\newcommand{\ProtocolA}{\ensuremath{\mathit{ProtocolA}}}
\newcommand{\ProtocolB}{\ensuremath{\mathit{ProtocolB}}}

\newcommand{\true}{\ensuremath{\mathit{true}}}
\newcommand{\false}{\ensuremath{\mathit{false}}}

         \def\cH{\ensuremath{{\mathcal H}}}
         \def\cL{\ensuremath{{\mathcal L}}}
         \def\cP{\ensuremath{{\mathcal P}}}

\newcommand{\set}[1]{\left\{ #1 \right\}}

\newcommand{\figlabel}[1]{\label{figure:#1}}
\newcommand{\nakedfigref}[1]{\ref{figure:#1}}
\newcommand{\figref}[1]{Figure~\nakedfigref{#1}}

\newcommand{\thmlabel}[1]{\label{thm:#1}}
\newcommand{\nakedthmref}[1]{\ref{thm:#1}}
\newcommand{\thmref}[1]{Theorem~\nakedthmref{#1}}

\newcommand{\seclabel}[1]{\label{sec:#1}}
\newcommand{\nakedsecref}[1]{\ref{sec:#1}}
\newcommand{\secref}[1]{Section~\nakedsecref{#1}}

\algdef{SE}% flags used internally to indicate we're defining a new block statement
[Class]% new block type, not to be confused with loops or if-statements
{Class}% "\Struct{name}" will indicate the start of the struct declaration
{EndClass}% "\EndStruct" ends the block indent
[1]% There is one argument, which is the name of the data structure
{\textbf{Class} \textsc{#1}}% typesetting of the start of a struct
{\textbf{end Class}}% typesetting the end of the struct

\newcommand{\examplei}{Example~\hyperlink{exp:e1}{I}\xspace}
\newcommand{\exampleii}{Example~\hyperlink{exp:e2}{II}\xspace}

\graphicspath{{./pictures/}}

\title{Invited Paper: Fault-tolerant and Expressive Cross-Chain Swaps}

\author{Yingjie Xue}
\email{yingjie_xue@brown.edu}
\affiliation{%
  \institution{Brown University}
  \city{Providence. RI}         % ACM template requires city and
  \country{United States}   % country for affiliation
}

\author{Di Jin}
\email{di_jin@brown.edu}
\affiliation{%
  \institution{Brown University}
  \city{Providence. RI}         % ACM template requires city and
  \country{United States}   % country for affiliation
}
\author{Maurice Herlihy}
\email{maurice.herlihy@gmail.com}
\affiliation{%
  \institution{Brown University}
  \city{Providence. RI}         % ACM template requires city and
  \country{United States}   % country for affiliation
}

\begin{abstract}
Cross-chain swaps enable exchange of different assets that reside on different blockchains. Several protocols have been proposed for atomic cross-chain swaps. However, those protocols are not fault-tolerant, in the sense that if any party deviates, no asset transfer can happen. In this paper, we propose two alternative protocols for structuring composable and robust cross-chain swaps. Participants can propose multiple swaps simultaneously and then complete a subset of those swaps according to their needs. Their needs are expressed as predicates which capture acceptable payoff of each participant. Our proposed protocols are thus more expressive due to the introduction of predicates. The proposed protocols are fault-tolerant since, even if some participants deviate, those predicates can still be satisfied, and conforming parties can complete an acceptable set of swaps.
\end{abstract}

\begin{document}
\maketitle

\section{Introduction}
\seclabel{intro}
Imagine a world where financial transactions span multiple independent ledgers,
each managing a different token.
Alice's salary is paid in \emph{apricot} tokens, managed on the Apricot ledger,
but she pays her utility bill using \emph{banana} tokens,
managed on the Banana ledger.
The apricot-to-banana exchange rate is volatile,
so Alice waits until the first day of each month to buy the banana tokens she needs.

Because Alice does not trust centralized exchanges,
she sets up a recurring trust-free atomic swap with some willing counterparty.
Suppose Alice and Bob agree to swap some of Alice's apricot tokens for Bob's banana tokens
using a well-known atomic swap protocol~\cite{tiernolan,Herlihy2018}
which require both Alice and Bob to place their tokens in escrow.
Any such protocol is \emph{safe} for Alice in the narrow sense that
if Bob deviates from the protocol,
perhaps by abandoning the swap in the middle,
then Alice gets her escrowed tokens back (minus fees).
Alice still pays a price because she regains access to her
escrowed tokens only after a substantial delay.
Alice must then attempt another swap with another counterparty,
exposing her to the same risk of inconvenience and delay.
Roughly speaking,
if Alice eventually succeeds on her $k^\text{th}$ attempt,
where each failed swap releases her escrowed tokens only after $t$ hours,
then Alice spends about $(k-1)t$ hours acquiring her banana tokens.

Suppose instead that Alice sets up all $k$ swaps together,
and tentatively executes those swaps in parallel.
Some swaps (tentatively) succeed and the rest fail.
Alice chooses to commit one of the successful swaps
(perhaps the one with the best exchange rate),
and aborts the rest.
As long as one tentative swap succeeds,
Alice acquires her banana tokens with a worst-case delay of $t$, not $(k-1)t$.

Prior cross-chain swap protocols are \emph{atomic} (all-or-nothing),
but not \emph{robust}:
any component failure typically causes the entire exchange to abort,
undoing any tentative changes.
Alternative paths can be explored only sequentially, not in parallel.
This paper proposes novel cross-chain swap protocols that allow parties
to explore multiple complex trades in parallel,
and to select some satisfactory subset of those trades to be completed.

Devising robust cross-chain swap protocols presents non-obvious challenges.
In the apricot-to-banana token swap example,
only Alice sets up alternative parallel swaps.
What if Bob, too, wants alternatives?
(Perhaps he buys banana tokens from Xerxes, or else from Zoe,
and then resells them to Alice.)
Can Bob and Alice's parallel swap alternatives compose in a well-defined way?
What if their choices interfere?

Prior cross-chain swap protocols
\cite{Herlihy2018,herlihy2021cross,decred2,shadab2020cross,zie2019extending}
typically ask participants to escrow their assets.
If all goes well,
the escrowed assets are redeemed by their new owners (the swap \emph{commits}),
but if something goes wrong,
the escrowed assets are refunded to their original owners (the swap \emph{aborts}).
The ``all-or-nothing'' nature of these protocols prevents a party
from redeeming only a subset of assets.
(Some payment channels~\cite{bagaria2020boomerang,rahimpour2021spear}
do support robustness though redundant payments,
but these are limited to one-way payments, not support cyclic transfers as in swaps.)

In this paper,
we propose two alternative protocols–\emph{ProtocolA}  and \emph{ProtocolB}, for structuring composable and robust cross-chain swaps.
These protocols make different trade-offs.
\begin{itemize}
\item
  We propose a framework where participants can set up multiple tentative
  swaps and commit only a subset.
  Participants express their requirements as predicates.
  The overall exchange is feasible if all parties' predicates are simultaneously
  \emph{satisfiable}.

\item
  We translate the predicates to a system of locks controlled by hashes~\cite{tiernolan}.

\item
  \emph{ProtocolA} has fast best-case completion time,
  but requires high \emph{collateral}:
  each party must fund a separate escrow for each tentative swap,
  even though some swaps will abort.

\item
  \emph{ProtocolB} has slower best-case completion time,
  but requires lower \emph{collateral}:
  the same escrow can be used in multiple alternative swaps.
\end{itemize}

This paper is organized as follows.
\secref{model} describes the model, \secref{motivation} elaborates our motivation,
\secref{preliminaries} introduces preliminaries for our proposed protocols.
\secref{problemOverview} describes challenges and some building blocks in proposed protocols.
\emph{ProtocolA} and \emph{ProtocolB} appear in \secref{protocolA} and \secref{protocolB}
respectively.
\secref{properties} analyzes the protocols' security and efficiency.
\secref{related} summarizes related work,
and \secref{remarks} considers future directions.

\section{Model of Computation}
\seclabel{model}
Although our problem is motivated by cross-chain asset exchanges,
nothing in our protocols depends on specific blockchain technologies,
or even whether ledgers are implemented as blockchains or some other kind of tamper-proof data store.
The fundamental problem addressed here is how to conduct fault-tolerant and safe commerce among
mutually-distrusting parties whose assets reside on multiple ledgers.

\subsection{Ledgers and Contracts}

A \emph{blockchain} is a ledger (or database) that tracks ownership of \emph{assets}.
A blockchain is tamper-proof,
meaning that it can be trusted to process transactions correctly and store data reliably.
We assume blockchains supports \emph{smart contracts} (or \emph{contracts}).
A smart contract is a program residing on the blockchain that can own
and transfer assets.
Contract code and state are public.
A contract is deterministic since it is executed multiple times for reliability.
A contract can read and write data on the same blockchain where it resides,
and it can invoke functions exported by other contracts on the same blockchain,
but it cannot directly access data or contracts on other blockchains.
A \emph{party} is a blockchain client,
such as a person or an organization.
A party can send transactions to be executed on the blockchain.
When we say \emph{transactions}, we mean transactions that happen on a single blockchain, including asset transfers, smart contract initialization,
and calling smart contract functions.

\subsection{Communication Model}\seclabel{communicationmodel}

We assume a \emph{synchronous} communication model
where there is a known upper bound $\Delta$ on the time needed for a transaction
issued by one party to included and confirmed \footnote{The mining process to include a transaction in a block and the confirmation of the block take a non-trivial time, see more in~\cite{poon2016bitcoin}. Local computation is considered instantaneous.}in a blockchain and to become visible to others.

\subsection{Fault Model}\seclabel{threatmodel}

As mentioned, blockchains are assumed to be tamper-proof,
and calls to smart contract functions are executed correctly.
We rule out attacks on blockchain itself, for example, denial-of-service attacks.
Parties can be \emph{Byzantine},
departing arbitrarily from any agreed-upon protocol.
We do not assume Byzantine parties are rational:
they may act against their own self-interest.
Because smart contracts typically reject ill-formed transactions,
Byzantine parties are limited in how they can misbehave.
\subsection{Cryptographic Assumptions}\seclabel{cryptomodel}
We make standard cryptographic assumptions. 
We assume a computationally bounded polynomial-time adversary. The hash function in our scheme is collision-resistant.  Each party is equipped a public key and a private key, and the public keys are known to all. Participants use standard public key algorithms, e.g. ECDSA \cite{johnson2001elliptic}, to sign their transactions so that they cannot be forged. We use $sig(m,u)$ to denote the signature generated by a party $u$ where he/she signs a message $m$ using his/her secret key.

\section{Motivation}
\seclabel{motivation}
\hypertarget{exp:e1}{} Suppose Alice owns $\Xcoin{}$s and she wants to buy an NFT from Bob. However, Bob only accepts payment in $\Ycoin$s. Alice finds an intermediate party, Carol, who accepts her $\Xcoin{}$s and in exchange pays $\Ycoin$s to Bob. Since Alice does not want to hold $\Ycoin$s if the trade fails, the three of them need to swap their assets atomically: Alice pays Carol $\Xcoin{}$s, Carols pays Bob $\Ycoin$s, and Bob sends the NFT to Alice (Shown in \figref{example1}, we call this \examplei).
\begin{figure}
\begin{subfigure}{0.49\columnwidth}
\includegraphics[width=\textwidth, height=6cm]{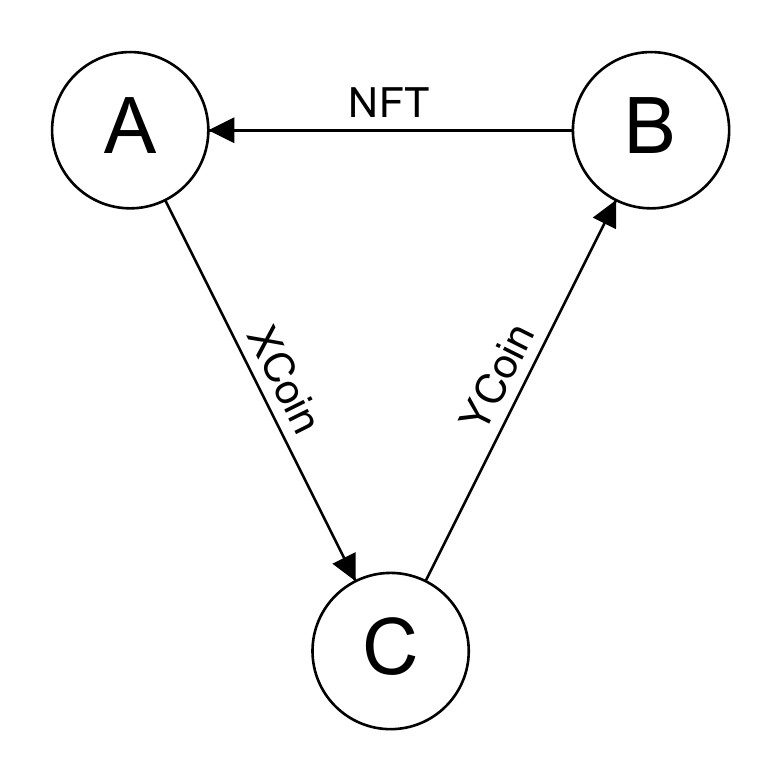}
\caption{\examplei: Trade With Carol}
\figlabel{example1}
\end{subfigure}
\hfill
\begin{subfigure}{0.49\columnwidth}
\includegraphics[width=\textwidth, height=6cm]{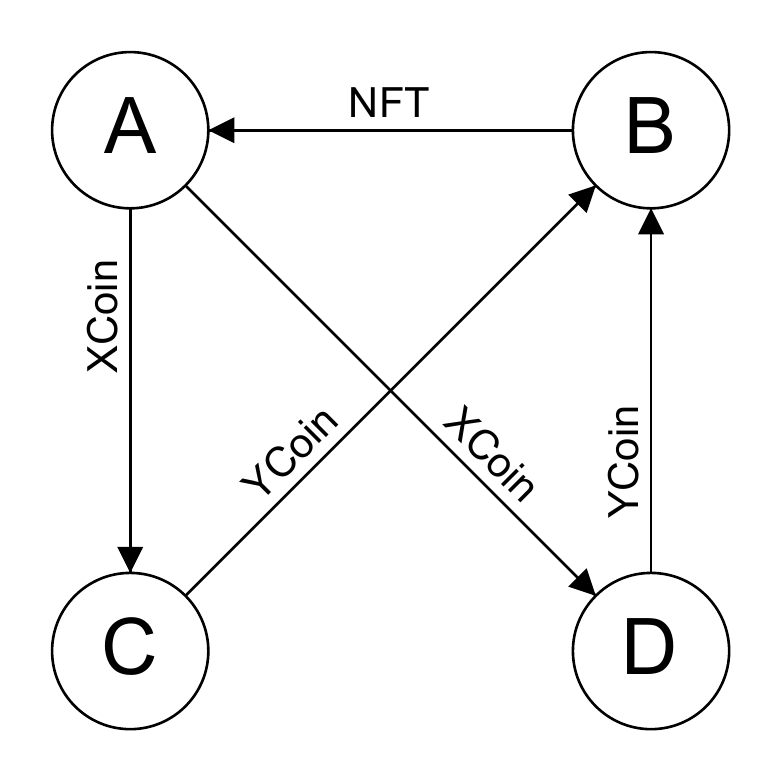}
\caption{\exampleii: Trade With Carol and David}
\figlabel{example2}
\end{subfigure}
%\vspace{-1.9em}
\caption{Two Motivational Examples}
\end{figure}

Atomic swap protocols \cite{Herlihy2018} are designed to handle this situation. They execute the tentative asset transfers with guaranteed atomicity: either all asset transfers happen, or no transfer happens.

\hypertarget{exp:e2}{} Suppose that Carol is not that reliable and she crashes with 50\% probability, but Alice and Bob want the trade to happen in a timely manner (the market might be very volatile). Alice and Bob can mitigate such problem by finding more trading partners (David) to boost their probability of success (\exampleii{} in \figref{example2}). As long as one of the trading partners is responsive, the trade can succeed.

We call what a participant wants to achieve in a trade their \textit{motive}, and a set of transactions that satisfies everyone's motive a \textit{feasible swap}. In \examplei, Alice's motive is to get the NFT from Bob while paying at most one $\Xcoin{}$ to Carol (she would be perfectly happy is she could get away with not paying), and in \exampleii, Alice's motive is to get the NFT from Bob, and paying at most one $\Xcoin{}$ to either Carol or David. Bob has similar motives in \exampleii. Carol and David have the same motive, either not doing any transaction or trading a $\Ycoin{}$ for an $\Xcoin{}$. As a result, in \examplei there is one feasible swap, but in \exampleii, there are two distinct feasible swaps: Alice pays P $\Xcoin{}$s, P pays Bob $\Ycoin$s, and Bob sends the NFT to Alice, where P is either Carol or David.

There is no easy way to run an atomic swap protocol to achieve the desired outcome: one of the two feasible swaps being completed. Intuitively, trying each feasible swap sequentially will take too much time, but running them in parallel requires resource sharing between different instances of the protocol, which is non-trivial.

\section{Preliminaries and Definitions}
\seclabel{preliminaries}
This section describes an existing atomic swap protocol, which lays the foundation of our proposed approaches. We start with terminologies in atomic swaps. The notation and terminology needed to define robust cross-chain swaps are also given in this section. 
\subsection{Directed Graphs}
An atomic cross-chain swap is represented by a \emph{directed graph}. A directed graph(or \emph{digraph}) $G=(V,A)$
is a set of \emph{vertices} $V$ and a set of \emph{arcs} 
$A \subseteq V \times V$.
Each vertex represents a party,
and each arc is labeled with an asset.
A \emph{path} from $u$ to $v$ in $G$ is a sequence of arcs
$(v_0,v_1), \ldots, (v_{k-1},v_k)$ where $u=v_0$ and $v=v_k$,
and each arc $(v_i,v_{i+1}) \in A$.
A digraph is \emph{strongly connected} if there is a path from
any vertex to any other vertex.

An arc $(u,v)$ represents a proposed asset transfer from 
party $u$ to party $v$.
All digraphs considered here are strongly connected,
implying that all transfers are (perhaps indirect) exchanges.
In real life, some exchanges may have off-chain components.
For example, if Alice uses a token to pay for a sandwich,
we would formally represent the off-chain food transfer as a ``virtual'' token
transferred from the restaurant to Alice.

For brevity, when discussing a swap defined by a digraph,
we use the terms \emph{vertex} and \emph{party} interchangeably.
We use "\emph{arc} $(a,b)$" as shorthand for "proposed transfer from party $a$ to party $b$",
and we say an arc is \emph{triggered} if that proposed transfer takes place.

\subsection{Hashlocks, Hashkeys and Hashlock Circuits}

Generally speaking, an atomic swap works in the following way: an asset is first escrowed and protected by locks, and then the asset is transferred to the recipient upon unlock. Here we describe the lock mechanism used in the atomic swap protocol described later.

Let $Hash(.)$ be a collision-resistant hash function. A pair of value $s$ and $h$, where $s$ is a secret random value, and $h=Hash(s)$, forms a ``lock and key'' structure:
 a lock $h$ can be unlocked when the key $s$ such that $h=Hash(s)$ is shown. A  \emph{hashlock} is such a lock, augmented with more information in the \emph{hashkey} to keep track of the propagation of the secret $s$ on the graph.
 
 Formally, a \emph{hashlock} is a hash value $h$. The structure to unlock hashlocks is called a \emph{hashkey}.  Given a digraph $G=(V,A)$, a \emph{hashkey} on an arc $(u,v)$ is a triple $(s,p,\sigma)$,
where $s$ is a randomly-chosen value called a \emph{secret},
$p$ is a path $(u_0,...,u_k)$ from $u_0$ to $u_k$ in the graph where $u_0=v$,
$u_k$ is the party who chose $s$. $\sigma$ is called a \emph{path signature}, where each party in the path $p$ provides a signature. Recall that  $sig(m,u)$ denotes the signature of a party $u$ signing a message $m$ using his/her secret key. $\sigma$ is a signature composed recursively by each party $u_j$ in the path signs the signature of $u_{j+1}$ as a message using their secret key. More formally,
\begin{equation}
\label{eqn:sigma}
\sigma =sig(sig(...sig(sig(s,u_k),u_{k-1})...,u_1),u_0)
\end{equation}
  A hashkey $(s,p,\sigma)$ has its lifespan. The exact definition of the lifespan and more details can be found in the original document ~\cite{Herlihy2018}. In short, the hashkey $(s,p,\sigma)$'s lifespan is linear to the length of the path: it times out at $|p|\Delta$, where $\Delta$ is the upper bound of message delay defined in \secref{communicationmodel} \footnote{More precisely, it times out at $(MaxPathLength(G)+|p|)\Delta$ after the protocol starts. Here we focus on brevity other than precision}.

We say a hashkey \emph{matches} a hashlock if it can \emph{unlock} the hashlock. On any arc $(u,v)$, its hashlock $h$ can be \emph{unlocked} by a hashkey $(s,p,\sigma)$ if and only if all of the following condition holds. (1) $h=Hash(s)$,(2) $p$ is a path from $v$ to the party who chose $s$. (3) $\sigma$ is a valid path signature for $s$ and $p$ constructed as Eq.\ref{eqn:sigma}. (4) The hashkey does not time out. Roughly speaking, the hashkey structure guarantees that, if a hashlock $h$ on an outgoing arc from $u$ is unlocked, then $h$ on all arcs entering $u$ can be unlocked.

A \emph{hashlock circuit} $C(u,v)$ for an arc $(u,v)$ is a formula linking hashlocks
on that arc via operators $\vee$, $\wedge$ and $\neg$.

\subsection{An Atomic Cross-Chain Swap Protocol}
\seclabel{multiparty_atomic_swap}

Here we describe a protocol ~\cite{Herlihy2018} that is atomic, but not robust.
It does not support alternative swaps,
but it does provide a starting point for developing robust protocols.

The swap is represented as a strongly-connected directed graph $G=(V,A)$,
where each arc represents a proposed asset transfer. Each transfer along an arc is controlled by a contract. We say an asset is escrowed on an arc when the owner forfeits the asset's control to the contract on the arc.

The protocol starts with a \emph{feedback vertex set} (FVS),
a set of vertices whose removal leaves the graph without cycles.
The vertices in the FVS are called \emph{leaders}, the rest \emph{followers}. 
At the beginning, each leader $l_i$ chooses a random secret $s_i$ and construct the hashlock $h_i=Hash(s_i)$. After the hashlocks are distributed among the leaders, the protocol can start running on the blockchains. Each contract/arc is associated with a hashlock circuit $C=\bigwedge_{l_i\in FVS}h_i$ to protect the escrowed asset. When a hashlock is unlocked, it evaluates to \true{} and when the hashlock circuit evaluates to \true{}, the asset is redeemed by the proposed recipient. We call an arc is \emph{triggered} when the asset is redeemed.

Overview of the protocol is shown in \figref{atomic_swap_protocol}.
The protocol consists of two phases: \emph{Escrow phase} and \emph{Redeem phase}. Note that transactions sent to blockchains can included on blockchains and observed by others within $\Delta$.

In the \emph{Escrow phase}, each leader escrows their assets on all outgoing arcs (\textbf{Leader Step a}). 
Then the leaders start waiting for their incoming arcs before they enter redeem phase (\textbf{Leader Step b}).
Followers first wait until all the incoming arcs are escrowed (\textbf{Follower Step a}), then they escrow their assets on the outgoing arcs (\textbf{Follower Step b}).

The total amount of assets by a party escrowed is called the party's \emph{collateral}. During the \emph{Escrow phase} if any expected escrow does not arrive for an extended period the party should just abort the protocol.

In the \emph{Redeem phase}, hashkeys are sent to contracts that manage their assets to unlock hashlocks. For a leader $l_i$, if it has observed that all its incoming assets are escrowed, it constructs a hashkey $(s_i, l_i, sig(s_i, l_i))$ and sends it to corresponding contracts representing all its incoming arcs (\textbf{Leader Step c}). The hashlock $h_i$ associated with the contracts can be unlocked upon receiving  $(s_i, l_i, sig(s_i, l_i))$. Then both leaders and followers can propagate the secrets (\textbf{Leader Step d} and \textbf{Follower Step c}) using the hashkey structure. For a party $u$, when a hashlock on its outgoing arcs is unlocked by a hashkey $(s, p, \sigma)$, it can construct and send the hashkey $(s_i, u|| p, sig(\sigma, u))$ to all contracts managing its incoming arcs to unlock the same hashlock. 

The hashkey mechanism guarantees that if any party observes a matching hashkey sent to its outgoing arc, there is one more $\Delta$ allowing it to send a new matching hashkey to all of its incoming arcs. If all secrets are propagated correctly, all hashlocks are unlocked, and all the assets will be redeemed.

There is an additional time-out structure that ensures that assets cannot be
escrowed forever, which is described in the original document~\cite{Herlihy2018}. In short, if a hashlock cannot be unlocked anymore, the asset will be refunded.
.

In summary, the atomic swap protocol satisfies liveness:
if all parties conform, all asset transfers happen.
It also satisfies safety: a conforming party $u$'s assets are all refunded if $u$
does not receive all its incoming assets. We call atomic swaps as \emph{all-or-nothing} swaps because of those properties.

\begin{figure}[!htp]
    \centering
    \includegraphics[width=\columnwidth]{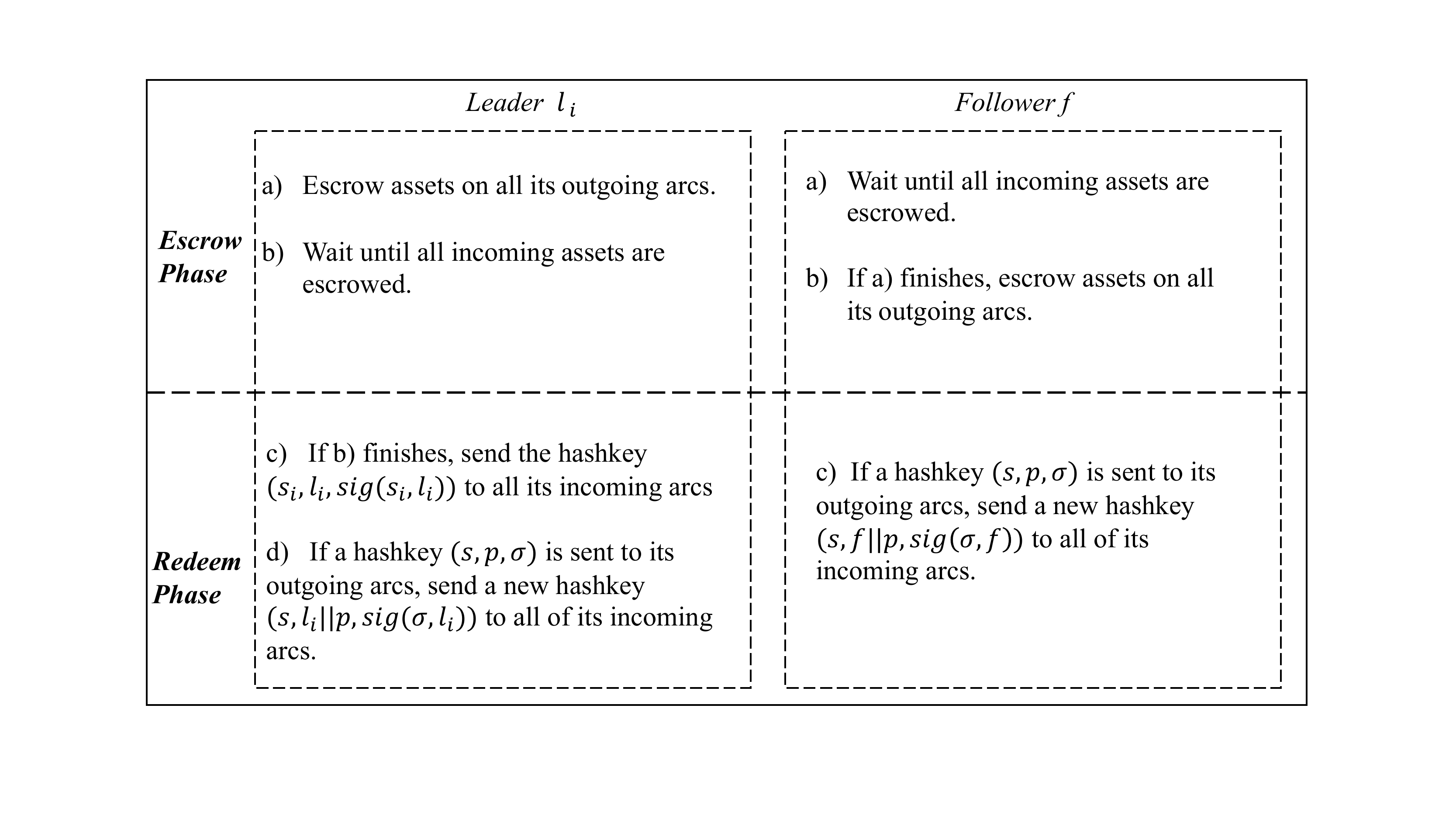}
\caption{The Atomic Swap Protocol}
     \figlabel{atomic_swap_protocol}
\end{figure}
\subsection{Predicates}
\seclabel{predicates}
The directed graph formalism suffices to describe 
\emph{all-or-nothing} swaps,
where success means all proposed transfers take place.
We now define a way to describe swaps where parties also accept outcomes where only a subset of the proposed swaps take place.

We first define a boolean variable $v$ on each arc. $v$ is \true{} if the transfer it represents takes place, and \false{} otherwise. An acceptable outcome for a party $x$ can be represented with a boolean predicate of its incoming and outgoing arcs.
Suppose party $x$ has incoming arcs $i_1$, $i_2$, outgoing arcs $o_1$, $o_2$, and $x$ wants to exchange $o_1$ with $i_1$, or exchange $o_2$ with $i_2$. Naively one would construct the predicate as $(o_1 \wedge i_1) \vee (o_2 \wedge i_2)$. Looking at it closely, this predicate definition has two problems. Firstly, it does not capture safety. When $o_1 \wedge i_1$ is \true{}, this predicate will be \true{} even if $o_2$ is \true{}, but this would mean that $x$ is paying both $o_1$ and $o_2$, i.e. $x$ overpays. Secondly, it does not allow $x$ to accept greedy outcome. If $i_1$ is \true{}, and all other arcs are \false{},  it is perfectly acceptable to $x$, but our predicate evaluates \false{} to this situation.

To capture their expected exchanges (liveness), safety requirements, and allow greediness,
\begin{comment}
Naturally, a party $x$ expects to receive some payments from other parties, and in exchange, they pay others. This gives us an income predicate $I_x$ for income they would like to receive, and an outgoing predicate  $O_x$ which characterizes acceptable outgoing payment. Easily said, how do $I_x$ and $O_x$ look exactly? A party $x$ may couple some incomes with some outgoing payments, for example, getting asset $a_1$ with paying one of the assets $b_1$ and $b_2$ (i.e. predicate $p_1=a_1\wedge \neg (b_1\wedge b_2))$, getting $a_2$ with paying one of the assets $c_1$ and $c_2$ (i.e. predicate $p_2=a_2\wedge \neg (c_1\wedge c_2))$, and getting $a_3$ or $a_4$ with paying $d_1$ (i.e. predicate $p_3=(a_3\vee a_4)\wedge (d_1 \vee \neg d_1))$. We see that there may be multiple income and outgoing payments, each combination has a predicate
$p$ (i.e. $p_1$, $p_2$, and $p_3$). Furthermore, in this example, the party $x$ may accept two of those exchanges completed. How do we express that? If we express as 2 of $p_1$, $p_2$ and $p_3$ being \true{}, then if one of them is not \true{}, then safety maybe broken, e.g. getting $a_1$ and pays both $b_1$ and $b_2$. Thus, we need a more delicate way to express their acceptable payoffs.
\end{comment}
each party has \emph{liveness} and \emph{safety} requirements,
characterizing acceptable outcomes in the (partial) success and failure cases respectively. First, we consider safety requirements. For each possible outgoing asset, there is an income predicate associated with it. That captures the payoff that, if this party pays this asset, what should they get at least to be safe. Note that, a participant may want some exchanges to happen atomically, say does not pay an asset unless getting a bundle of other assets even though get a subset of them already gives him/her a better payoff. This is also characterized as a safety requirement. The safety requirement for party $x$ is given by a predicate $S_x$. 

In \examplei, we use the arc "$(a,c)$" as shorthand for
``the asset labeling arc $(a,c)$ is transferred from party $a$ to party $c$''. Alice's safety predicate is
\begin{equation*}
  S_a := (a,c) \implies (b,a),
\end{equation*}
meaning that if Alice transfers her assets to Carol (``$(a,c)$'')
then Bob transfers his assets to her (``$(b,a)$'').
Importantly, Alice can be greedy.
Her predicate is satisfied if she gets something for nothing:
that is, if Bob pays her but she somehow does not pay Carol.
This predicate is also satisfied if no payments are made,
an outcome that Alice may not prefer,
but considers acceptable because it leaves her no worse off. The no payment scenario incentives Alice to try alternatives.

Formally, for every asset $\gamma$ that a party $x$ may pay, there is an income predicate $I_x^\gamma$. In addition, there is a predicate $O_x$ over those outgoing payments to make sure $x$ does not overpay, 
for example, the predicate may require that at most two of three outgoing assets are paid.

In \exampleii, Alice's predicate is

\begin{equation*}
\begin{split}
  S_a := &\left((a,c) \implies (b,a)\right) \\
  &\wedge \left( (a,d) \implies (b,a)\right)\\
  &\wedge \neg \left((a,c) \wedge (a,d)\right).
  \end{split}
\end{equation*}
The first two clauses say that if Alice pays either Carol or David,
then she gets Bob's NFT in exchange,
and the third clause says that Alice does not want to pay both.

In general, if party $x$ has potential exchanges with parties $u_1, \ldots, u_k$,
then $x$'s safety predicate has the form:
\begin{equation*}
  S_x:= \left( \bigwedge_{i=1}^k ((x,u_i) \implies I_{x}^{(x,u_i)}) \right) \wedge O_x,
\end{equation*}
where each implication clause states that if $x$ pays $u_i$ it gets the agreed-upon amount in return,
and the final clause limits how many of the outgoing payments' clauses $(x,u_i)$ can be \true{}.
For example $O_x$ might say ``no more than one of the $k$ transfers can occur'', or ``no more than $m$'', or ``at least $m$'', and so on.

Next, we consider liveness requirements.
The liveness requirement for party $x$ is given by a predicate $L_x$. First, the liveness predicate contains the safety predicate since this is a property that should always hold. In addition to safety, liveness characterizes that something good should happen. That means, one of the income predicates $x$ specified previously in safety requirements should be \true{}. That is,

\begin{equation*}
L_x := S_x \wedge \left(\bigvee_{i=1}^k  I_{x}^{(x,u_i)}\right)
\end{equation*}

The predicate for each party $x$ is denoted as $P_x$, which consists of two predicates $L_x$ and $S_x$. Unless otherwise specified, $P_x$ means $S_x$, since $S_x$ should always be \true{}, while, reasonably, $L_x$ is \true{} only if a party completes an asset transfer. Thus, $L_x$ is implied implicitly in $P_x$, since if a party transfers an asset to someone, one of its income predicates in $S_x$ must be \true{}.

\subsection{Example}
\seclabel{example}
Suppose Alice would like to exchange 1 Xcoin for 1 Ycoin.
She sets up alternative trades with Bob and Carol,
but she is willing to trade with only one of them.
Bob expects to exchange 1 Ycoin for 1 Xcoin, with either Alice or Carol, but not both.
Carol is willing to trade with Alice, or Bob, or both.

\begin{figure}[!htp]
    \centering
    \includegraphics[width=.3\textwidth]{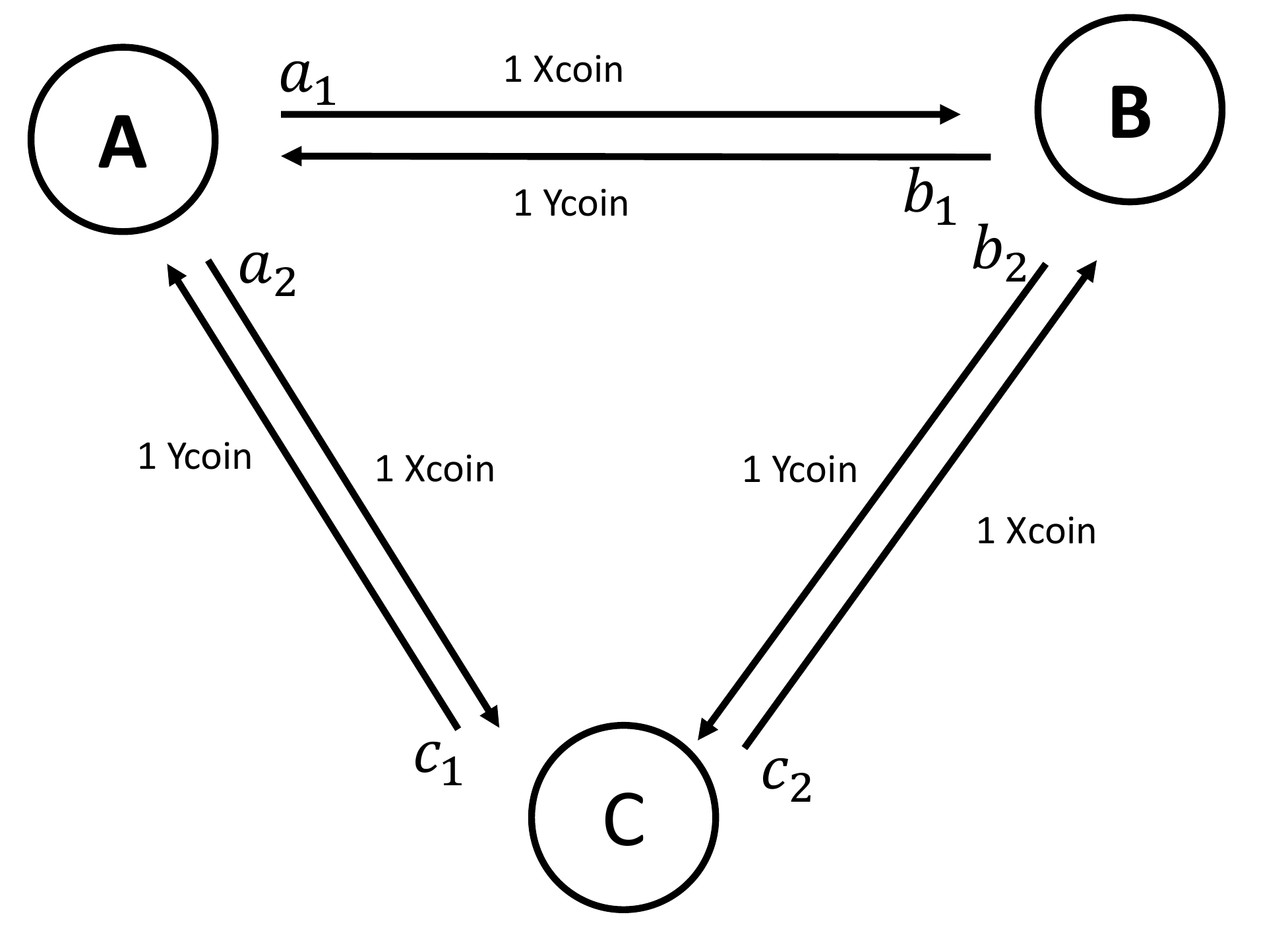}
\caption{Example for A Swap With Predicates}
     \figlabel{example}
\end{figure}
The predicates are as follows. For brevity, arcs are labeled as shown in \figref{example}.
(For example, $a_1$ is the arc from Alice to Bob.)
\begin{equation*}
  \begin{split}
    \text{Alice: } &P_a := (a_1 \to b_1) \wedge (a_2 \to c_1) \wedge \neg (a_1 \wedge a_2) \\
    \text{Bob: } &P_b := (b_1 \to a_1) \wedge (b_2 \to c_2) \wedge \neg (b_1 \wedge b_2) \\
    \text{Carol: } &P_c := (c_1 \to a_2) \wedge (c_2 \to b_2)
  \end{split}
\end{equation*}

\section{Roadmap and Building Blocks }
\seclabel{problemOverview}
This section describes a roadmap to devise a robust cross-chain swap protocol, the challenges, 
and the building blocks from which we construct our approaches.
\subsection{Roadmap}
A robust cross-chain swap can be described by a digraph $G=(V,A)$ accompanied with a set of predicates from all participants. Each arc in $G$ is a proposed asset transfer. An arc is set to \true{} if the asset transfer happens. A predicate is satisfied if it evaluates to \true{}. Is there a trade that satisfies every party's predicate?
Whether such a trade exists is the \emph{satisfiability} problem.
If not, then the proposed trade is infeasible. 

We should first find solutions that satisfy all parties' predicates. A \emph{solution} $s$ is an assignment that satisfies all parties' predicates, where each arc is assigned a Boolean. \footnote{The trade where no assets are transferred is acceptable to all parties,
but we exclude that solution as trivial.} For example,
here is one possible solution for the digraph shown in \figref{example}:
\begin{equation*}
\set{a_1 \mapsto \true, a_2 \mapsto \false, b_1 \mapsto \true, b_2 \mapsto \false, c_1 \mapsto \false, c_2 \mapsto \false}
\end{equation*}

After finding solutions, we can map them to feasible swaps. In each solution $s$,
we look at the set of arcs that $s$ assigns to \true{}, which forms a digraph denoted as $G_s$. The set of transfers in $G_s$, if executed atomically, will satisfy all parties. A digraph $G_s$ is a called feasible swap, or an alternative.

If there is more than one such feasible swap,
it is tempting to execute all the alternatives in parallel,
because some alternatives might fail. The following describes challenges that arise if we try to execute
the alternatives in parallel.

\subsection{Challenges}

\begin{enumerate}
\item
  Alternatives may conflict:
  in one solution, Alice pays Bob and not Carol, but in another, she pays Carol and not Bob.
  Completing both swaps would cause Alice to overpay.
  
\item
  Alternatives may charge twice for the same transfers:
  if there are two solutions where Alice pays Bob,
  then Alice has to escrow the same amount of assets twice.
Alice's collateral would exceed the value of the assets she trades away.
  
\item
  Trades are not independent when one alternative's digraph is a subgraph of the other.
  For example, \figref{connection} shows two solutions.
  In $s_1$, Alice and Carol trade only with one another, setting only $a_2, c_1$ to \true.
  Suppose Alice is the only leader.
  In $s_2$, Alice trades with Carol, and Carol with Bob,
  setting $a_2,c_1,b_2,c_2$ to \true.
  Suppose Alice and Bob are leaders.
  To complete both alternatives,
  Alice would have to create and release secrets for both $s_1,s_2$.
  She might not have an incentive to participate in $s_2$,
  since that alternative provides her no additional robustness.
 
\end{enumerate}
\begin{figure}
    \centering
     \includegraphics[width=.3\textwidth]{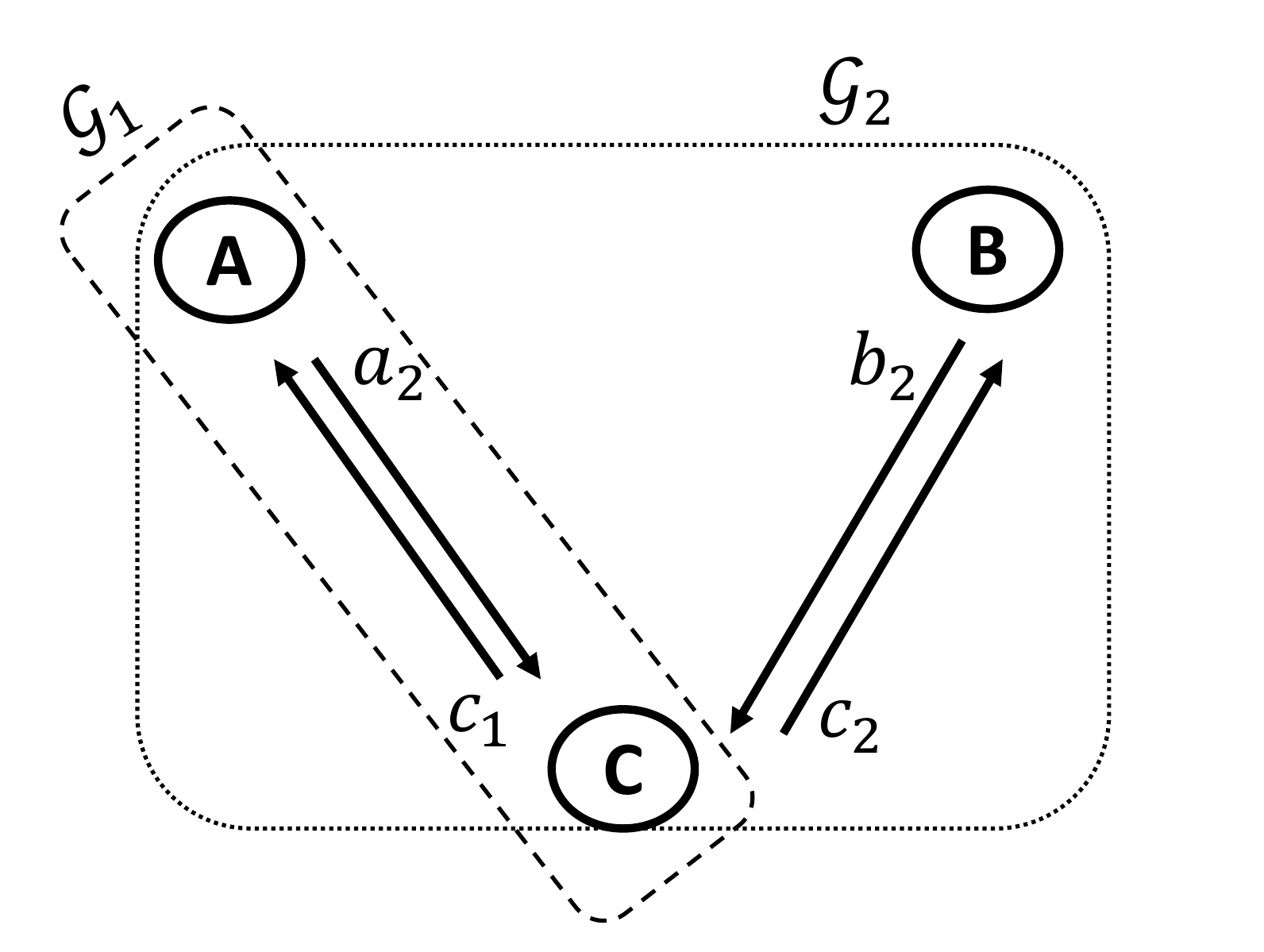}
    \caption{A solution is part of another solution}
    \figlabel{connection}
    \end{figure}

    \begin{figure}
   \centering
     \includegraphics[width=.3\textwidth]{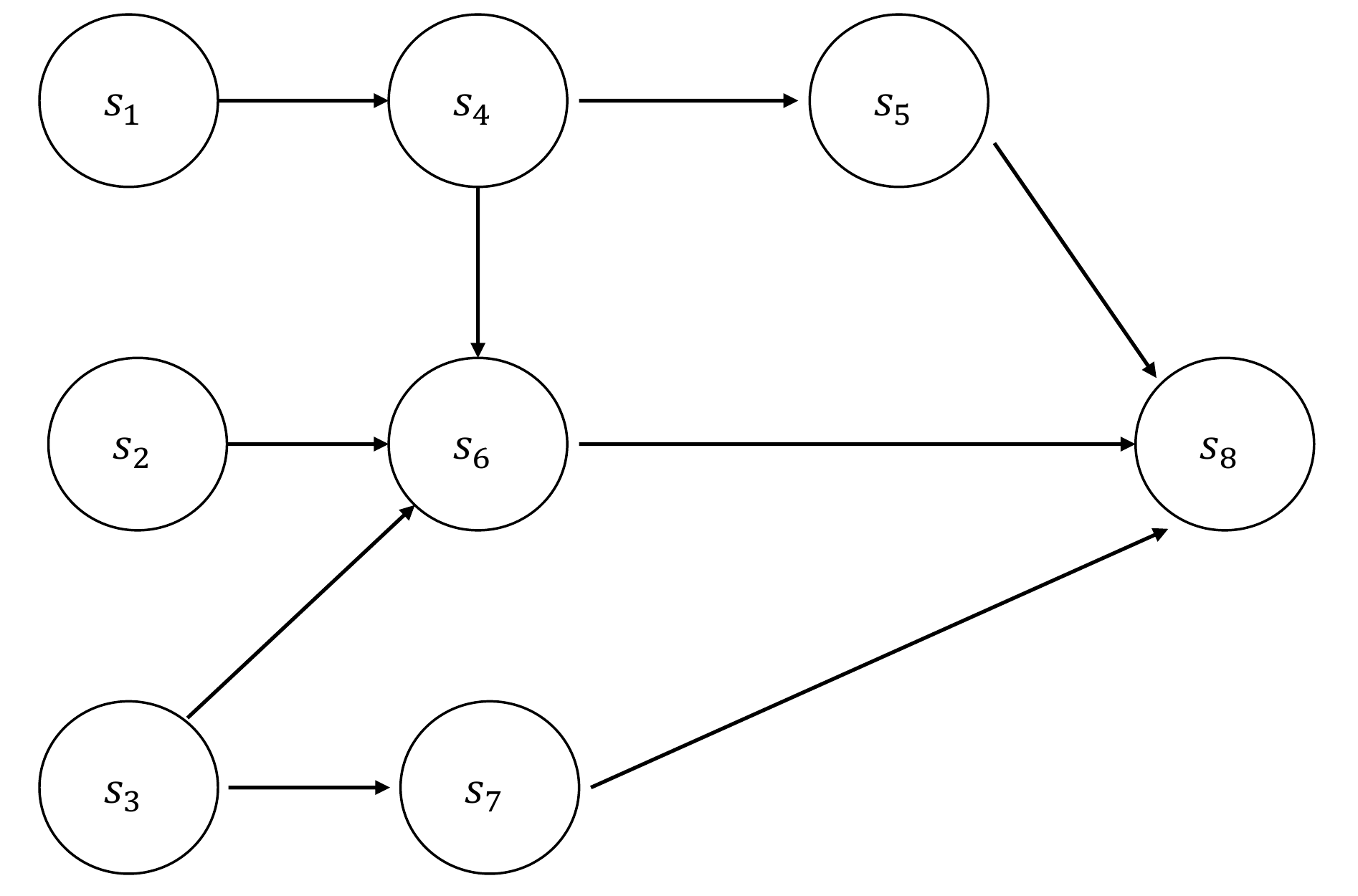}
    \caption{DAG Representation of Solutions}
     \figlabel{solution_DAG}
\end{figure}

\subsection{Building Blocks in Proposed Protocols}
To tackle the challenges,
we first describe some concepts and building blocks  for our proposed protocols.

\subsubsection{Mapping Arc Assignments to Swap Digraphs}
A robust cross-chain swap is described by a \emph{predicated directed graph} defined as
$(\cP,G)$, where $G$ is the digraph of all proposed transfers,
and $\cP$ is the set of all parties' requirements.
$\phi(\cP)$ denotes the conjunction of predicates $p\in \cP$:
\begin{equation*}
  \phi(\cP) = \bigwedge_{p \in P} p.
\end{equation*}

Given a digraph $G=(V,A)$,
an \emph{assignment} is a map $\alpha: A \to \set{\true,\false}$
that assigns a Boolean to each arc. For any assignment $\alpha$,
$\phi(\cP)(\alpha)$ denotes the value of $\phi(\cP)$ under assignment $\alpha$. A \emph{solution} $s$ is an assignment where $\phi(\cP)(s)=true$. 
A \emph{swap digraph} $G_{s}$  contains the set of arcs that $s$ assigns to \true{}, denoted as 
\begin{equation*}
G_s=\set{(a,b) \in A | (a,b)\mapsto \true \in s}.
\end{equation*}

Some $G_s$ is a subgraph of a larger graph $G_{s'}$. We define this relation as  inclusion $\subsetneq$.

\begin{definition}
\sloppy Inclusion $\subsetneq$. Consider solutions $s,s'$ where  $\phi(\cP)(s)=\phi(\cP)(s')=\true$.  $G_s=(V_s,A_s)$ and $G_{s'}=(V_{s'}, A_{s'})$. If $V_s \subset V_{s'}$ and  $A_s\subsetneq A_{s'}$, then we say $s \subsetneq s'$.
\end{definition} 

\subsubsection{Redundancy Providers}
Some parties may prepare redundant trades for fault tolerance
even though they do not intend to complete all of those trades.

\begin{definition}
\label{redundancy_provider}
  Let $(\cP,G)$ be a predicated directed graph, and $x$ a vertex in $G$.
  If there are solutions $s_1, s_2$,
  where $x\in V_{s_1}$ and $x\in V_{s_2}$,
  such that $\phi(\cP)(s_1)= \phi(\cP)(s_2)=\true$,
  but setting all the arcs in $G_{s_1}$ and $G_{s_2}$ to \true{} makes $P_x=\false$,
then $s_1$ and $s_2$ are called {\bf{conflicting}} solutions, and $x$ is called a {\bf{redundancy provider}}.
\end{definition} 

\subsubsection{Swap Schemes}
Next we describe a swap scheme derived from the atomic swap protocol ~\cite{Herlihy2018} with minor changes,
which lays the foundation for the robust protocols defined later.

A \emph{swap scheme} is a tuple $(G,\cH)$,
where $G$ is a digraph and $\cH$ the set of hashlocks on each arc.
The circuit $C=\wedge_{\forall h\in \cH} h$
denotes that all hashlocks in $\cH$ need be unlocked to trigger any arc.
Each arc has the same hashlock set $\cH$. In a swap scheme,
 a swap $(G,\cH)$ is executed in two sequential phases:
the \emph{Escrow Phase} (denoted as $\Swap.\Escrow(G,\cH)$),
and the \emph{Redeem Phase} (denoted as $\Swap.\Redeem(G,\cH)$).

Compared to the original swap protocol ~\cite{Herlihy2018} described in \secref{preliminaries}, the only difference in our swap scheme is that, the set $\cH$ in the original scheme is the set of hashlocks generated by leaders, while in ours, $\cH$ also contains a set of hashlocks generated by redundancy providers. As a result, in the \emph{Redeem Phase}, redundancy providers also need to send their hashkeys like leaders do.

The interface we defined in the swap scheme works as follows. 

\begin{itemize}
\item $\Swap.\Escrow(G,\cH)$
  \begin{itemize}
  \item
   The same as  the original swap protocol.
\end{itemize}
\item $\Swap.\Redeem(G,\cH)$
\begin{itemize}
\item If $x$ is a leader or a redundancy provider,
  send hashkeys as a leader in the original swap protocol.
    
\item
  Once any party $x$ receives a hashkey on their outgoing arc, it sends a new hashkey to their incoming arcs as in the original swap protocol.  
\end{itemize}
\end{itemize}

We intend to run different swap schemes in parallel. Here we define a maximal set of compatible schemes that can be completed.

\begin{definition}
  A maximal set of schemes. A set of schemes is maximal, if all schemes in the set are not conflicting, and there is no new swap schemes can be added to the set without having a conflict with some of existing schemes.
\end{definition}

\subsection{Two Protocols and a Trade-off}
We propose two novel protocols for composable and robust cross-chain swaps.
These protocols make different trade-offs:
one optimizes for \emph{time} at the cost of higher \emph{collateral},
and the other makes the opposite choice.

\ProtocolA{} prioritizes time over collateral.
In the best case, when all parties conform to the protocol and avoid delays,
the trade can be completed in a short time.
In the worst case, when all trades fail,
the parties' assets are refunded quickly.
The catch is that parties must provide higher collateral:
parties must create separate escrows for each alternative,
temporarily tying up more assets than are eventually traded.

\ProtocolB{} prioritizes collateral over time.
A party can use a single escrow for multiple alternatives.
The catch is that the trade takes longer to settle:
transfers require a hard timeout even when all parties conform to the protocol
and avoid unnecessary delay.
\ProtocolB{} also provides more fault-tolerance than \ProtocolA{}. See \secref{comparison} for more details.

\section{ProtocolA: Higher Collateral}
\seclabel{protocolA}
\subsection{Overview}

This section describes \ProtocolA,
a protocol that favors time over collateral.
Given a predicated digraph $(\cP,G)$,
we first find a set of solutions $s_1, \cdots s_k$
by identifying sets of arcs
such that executing trades on those sets satisfies $\phi(\cP)$.
Such solutions could be generated by
an \emph{all-SAT {solvers}}~\cite{toda2016implementing,yu2014all,todarepo},
though it may not be necessary to identify all solutions.
The parties run these solutions in parallel. The follow summarizes \ProtocolA. 
\begin{itemize}
\item
  To control conflicting trades,
  each redundancy provider uses distinct hashlocks in distinct solutions.
  In the \emph{Redeem Phase}, that party can choose which hashlocks to unlock.
  
\item
  A party can add additional hashlocks after an asset is escrowed,
  allowing overlapping arcs in different solutions,
  indexed by their hashlocks.
    
\item
  If $G_{s_1}\subsetneq G_{s_2}$, then the hashlocks on $G_{s_1}$ are reused in $G_{s_2}$.
\end{itemize}

\subsection{Detailed Construction}

First, the parties set up a mutually-agreeable trade,
and express their requirements in the form of predicates,
yielding a predicated digraph $(\cP, G)$. We assume the predicates are reasonable and if not, a party can be rejected to join in the trade.
A preliminary \emph{Market Clearing Phase} decides \emph{what swaps are feasible, and what hashlocks to use in each swap.}

\subsubsection{Market Clearing Phase}

\paragraph{Find solutions}
Given $(\cP, G)$,
the first step is to find a set of solutions acceptable to all parties,
perhaps by applying an all-SAT solver to $\phi(\cP)$,
yielding assignments $\alpha$ for which $\phi(\cP)(\alpha)$ evaluates to \true.
If we do not need all solutions,
we can stop after finding enough assignments.
Suppose we have found $k$ solutions $S= \set{s_i|\phi(\cP)(s_i)=\true, i\in[1,k]}$.
We rule out solutions that are not strongly connected,
since if the graph is not strongly connected,
some rational parties have incentive to deviate~\cite{Herlihy2018}.

\paragraph{Sort solutions}
Each digraph $G_{s_i}$ corresponds to a swap, and we construct schemes to execute the swap as atomic swap does. Since we plan to reuse hashlocks if some solutions $s$, $s'$ satisfies $s\subsetneq s'$, solutions are sorted by inclusion.
This can be done trivially by comparing each pair of solutions $(s_i,s_j)$.
We use a directed graph $T$ to depict their relation (\figref{solution_DAG}),
where an arc from $s_i$ to $s_j$ means $s_i\subsetneq s_j$.  $T$ is a directed acyclic graph (DAG) since $\subsetneq$ is not  reflexive.
In \figref{solution_DAG},
each node is a solution and each arc is a $\subsetneq$ relation.
For example, $s_1\subsetneq s_4\subsetneq s_5$.
If one solution, say $s_8$, is reachable from another solution, say $s_1$,
then $s_1\subsetneq s_8$.
(There is no need for a direct arc $(s_1,s_8)$ since inclusion can be inferred). Note that if there is an arc $(s,s')$ in $T$, then no $s''$ exist such that $s\subsetneq s'' \subsetneq s'$.
If one solution is not reachable from another, then those solutions are incomparable.

We call solutions that are not reachable by any other solutions \emph{root solutions}.
The solutions that they can reach directly are called their \emph{children}.
Solutions that do not reach other solutions are called \emph{leaves}. 
A path to a leaf node $v$ is denoted as $q=[v_0, \cdots,v]$,
where $v_0$ is a root node and $v$ is reachable from $v_0$.
In the graph $T$, all paths from all roots to reachable leaf nodes in the tree are denoted as $Q(T(S))$, where $T(S)$ is a DAG of solutions in the set S.

\paragraph{Assign hashlocks}
After sorting solutions in $S$,
we are ready to assign hashlocks to swap diagraph $G_s=(V_s,A_s)$ where $s\in S$.

We first assign hashlocks to the root solutions.
For any root solution $s$, if the corresponding $G_s$ is cyclic,
like in atomic swap, then we choose a feedback vertex set (FVS).The vertices in FVS are called \emph{leaders} $\cL$.
Although finding a minimum feedback vertex is NP-complete,
there exists an efficient 2-approximation~\cite{BeckerG1996}.
Recall that two exchanges are conflicting
if there exists a party $x$ such that $P_x=\false$ if both exchanges are completed.
We identify the set of redundancy providers $RP$ by checking whether a party $x$ is
involved in two conflicting solutions such that $P_x=\false$ if two conflicting exchanges are both completed.
The set of hashlock generators is $HG=RP \cup \cL$.
Each party $x\in HG$ generates a hash $h_{x}^{s}=Hash(\theta_{x}^s)$
meaning the hashlock is used for solution $s$
generated by a party $x$, and the secret is $\theta_{x}^{s}$.
For all arcs in solution $s$,
the set of hashlocks $\cH_s$ is $\cH_s=\{h_{x}^{s},
\forall x\in HG\}$ and the corresponding circuit is
$C_s=\bigwedge h, \forall h \in \cH_s$ . 

After assigning hashlocks for root solutions,
we move to their children. Their children will reuse the hashlocks from them.
Note that a root can have multiple children, and a child can have multiple parents.
For this reason,
a solution's hashlock may be used by multiple children,
and a solution may reuse hashlocks from multiple parents.
From each root node in the graph $T$,
we search all paths from the root to its children in the tree until the leaf.
A path is denoted as $q=[s_0, s_1,s_2,\cdots,s_k]$,
where each $s_i$ denotes a node and $s_0$ is a root node.
For ease of exposition,
we used $(G_s,\cH_{s,q})$ to mean that all arcs in $G_s$
are assigned hashlock set $\cH_{s,q}$ for solution $s$ in the path $q$.
For root solutions,
$\cH_{s,q}$ is the same for all $q$ since $q=s$.
For non-root node $s$,$\cH_{s,q}$ is different for different $q$.

Starting with a root solution $s_0$,
we assign hashlocks for solutions reachable from $s_0$.
Here we show how to assign hashlocks for solutions
in a path $q=[s_0,s_1,s_2,\cdots, s_k]$ starting from $s_0$.
For $i\in [1,k]$, assume $s_{i-1}$ is associated with hashlock set $\cH_{s_{i-1},q}$.
Then, for $s_i$, the hashlocks are set using the following steps.
\begin{itemize}
\item Compute $G_{s_i\setminus s_{i-1}}=A_{s_i}\setminus A_{s_{i-1}}$
  and find hashlock generators for $G_{s_i\setminus s_{i-1}}$. We also use $s_i\setminus s_{i-1}$ to denote $G_{s_i\setminus s_{i-1}}$ when there is no confusion.
  \begin{itemize}
  \item If  $G_{s_i\setminus s_{i-1}}$ is acyclic, then no leader is introduced.
  \item If $G_{s_i\setminus s_{i-1}}$ is cyclic,
    then new leaders in $G_{s_i\setminus s_{i-1}}$ are chosen.
    Denote by $\cL_{s_i\setminus s_{i-1}}$ the set of leaders in $G_{s_i\setminus s_{i-1}}$.
  \item If redundancy providers are introduced in $G_{s_i\setminus s_{i-1}}$,
    then redundancy providers are added.
    Let $RP_{s_i\setminus s_{i-1}}$ denote the redundancy providers in $G_{s_i\setminus s_{i-1}}$.
  \item New hashlock generators $HG_{s_i\setminus s_{i-1}}=\cL_{s_i\setminus s_{i-1}} \cup RP_{s_i\setminus s_{i-1}}$
    \end{itemize}
    \item Each party in $HG_{s_i\setminus s_{i-1}}$ generate a new hashlock. And the set of hashlocks generated is denoted as   $\cH_{s_i\setminus s_{i-1}}$. 
    \item The hashlocks for $G_{s_i} $ on the path $q\in Q(T(S))$ is $\cH_{s_{i},q} =\cH_{s_{i-1},q} \cup  \cH_{s_i\setminus s_{i-1}}$.
\end{itemize}

\subsubsection{Execute the protocol on chain}

After assigning hashlocks for solutions, each solution $s$ can be described as a set of swap schemes
$\Swap(G_s, H_{s,q}), \forall q \in Q(T(S))$. Those swap schemes can be executed in parallel on the chain. Denote the solutions output by all-SAT solvers as $S=\{s_1,\cdots, s_k\}$. 

For each party $x$, we first find all solutions involving it,
i.e. $x\in V_{s_i}$.
Then for each such solution,
it executes a separate swap scheme $\Swap(G_{s_i},
\cH_{s_i,q})$ for all  $q\in Q(T(S))$, in three phases
called \emph{escrow}, \emph{select} and \emph{redeem}.

\paragraph{Escrow Phase}
\sloppy Each party in $\Swap(G_{s_i}, \cH_{s_i,q})$ runs $\Swap.\Escrow(G_{s_i}, \cH_{s_i,q})$ in parallel. 
If an asset is already escrowed, it is not escrowed again.
Instead, the asset's circuit is updated with an OR gate:
$C_{s_i,q}\vee C_{\current}$. 
Suppose $C_{\current}$ means the current hashlock circuit on an arc. $C_{s_i,q}$ is the hashlock circuit for $\Swap(G_{s_i}, \cH_{s_i,q})$, which is the conjunction of all hashes in $\cH_{s_i,q}$.
Symmetrically, parties do not require incoming assets to be escrowed twice,
only that the hashlocks on those incoming assets are updated to $C_{s_i,q}\vee C_{\current}$. A party can update the hash circuit on its outgoing arc using OR gate since it adds more possibilities to be redeemed and does not affect the ability to redeem using the current hash circuits.

\paragraph{Select Phase}
After the escrow phase is finished,
the parties select which swap scheme should proceed, since some swap schemes are conflicting so that not all of them can be completed.
First, redundancy providers run an agreement procedure to decide which set of swap schemes
they would like to complete. The agreement procedure is not the main focus of the protocol, here we give an algorithm to reach an agreement without considering its efficiency.

A redundancy provider is randomly chosen as a \emph{proposer}.
That party proposes a maximal set of compatible schemes $S_c$
(which can be generated by a greedy algorithm)
from  all swap schemes where no party deviates in the escrow phase.
The rest of the redundancy providers are \emph{voters},
who vote whether to complete those schemes.
If the proposer is conforming, $S_c$ should be acceptable for all of them,
and conforming voters should all vote yes.
If some of voters do not vote yes,
then this scheme is removed from $S_c$.
Another round lets the proposer add schemes to $S_c$ to make sure $S_c$ is maximal.
The role of proposer can be replaced by a program shared among all parties,
which observes all escrows on the chain and deterministically
outputs a maximal set of schemes $S_c$.
Redundancy providers then vote on each scheme in this set $S_c$.
The search for $S_c$ ends when all redundancy providers vote yes on each proposed swap scheme,
or there are no new swap scheme to be added to $S_c$.
For fast settlement, this protocol can run on the side.
For example, once the escrow phase in a swap scheme $s_1$ is completed,
the protocol can start to decide whether to complete the swap.

\paragraph{Redeem Phase}
After $S_c$ is chosen by the redundancy providers,
the parties proceed as follows. We use a tuple $(s,q)$ to denote the swap scheme $\Swap(G_{s}, \cH_{s,q})$.
A leader or redundancy provider $x$ needs extra consideration before they proceed to the redeem phase for a scheme $(s,q)\in S_c$ because of the reuse of hashlocks. They proceed in  $\Swap(G_{s}, \cH_{s,q})$ only if
the hashlock $h$ that $x$ generated for $(s,q)$ satisfies:
for all $\Swap(G,\cH)$ where $h\in \cH$, 
$x$ receives all incoming escrows in $\Swap(G,\cH)$.
 Otherwise, it does not proceed. This requirement guarantees that, if a leader/redundancy provider releases a hashkey for a hashlock $h$, and any outgoing arc is triggered in any scheme $\Swap(G_s,\cH)$ who uses this $h$, then they can get assets from incoming arcs in this scheme, leaving them no worse off. See more analysis in \secref{proof}. If $x$ is not a leader or redundancy provider, no extra consideration is required before they proceed. For each selected swap scheme $\Swap(G_{s_i}, \cH_{s_i,q})$,
those who proceed run $Swap.\Redeem(G_{s_i}, \cH_{s_i,q})$.

An asset escrowed can be redeemed by a counterparty if its hashlock circuit
evaluates to \true, where we assign \true{} to hashlocks that have been unlocked,
and \emph{false} to the rest.
Recall that the circuit is composed of the disjunction of hashlock circuits
of each independent swap scheme. That means, if the circuit of any independent swap scheme evaluates to \true{}, a swap happens.

\section{ProtocolB: Lower Collateral}
\seclabel{protocolB}
Suppose Alice wants to exchange one apricot token for one banana token.
Using \ProtocolA,
Alice sets up tentative trades with Bob, Carol, and David.
She must escrow three 3 apricot tokens, one for each possible trade.
In the end, she will commit one trade, spending one token,
and cancel the rest, reclaiming the other two tokens.
Nevertheless,
she must have three tokens at hand to provide collateral for the alternative trades.

In this section, we describe \ProtocolB,
a protocol that allows Alice to provide collateral for all three alternatives with the same token.
The catch is that \ProtocolB{} requires a hard timeout to complete the trade.
In both the best and worst cases,
\ProtocolB{} takes time $4\Delta$ \footnote{Each swap takes 4 rounds, where $\Delta$ is the upper bound of message delay for a round.},
while \ProtocolA{} requires less time since participants can complete an alternative immediately after its \emph{escrow phase} is completed.
Note that even with a hard timeout,
\ProtocolB{} requires less time than attempting the alternative trades sequentially.

\subsection{Overview}
Here is a high-level sketch of \ProtocolB{}.

We focus on the difference between  \ProtocolA{} and \ProtocolB{} in the description.
\paragraph{Predicates reflect the reuse of assets.} To start, parties express their exchange requirements just as for  \ProtocolA{}. The difference between  \ProtocolA{}  and  \ProtocolB{} is that, in  \ProtocolA{}, each arc represents a unique asset, while in \ProtocolB{}, some arcs can represent the same asset, e.g. one token is reused on multiple arcs. To cater for that change, each participant provides an additional predicate: for different arcs that represent the same assets, at most one of them is assigned \true. Given the participants' new predicates, we find assignments to satisfy all the predicates.

\paragraph{Solutions are sorted by preferences.} Suppose there are $k$ solutions. We assign hashlocks as in \ProtocolA{} (the definition of redundancy providers change a bit, explained later). For an asset that has multiple different recipients, solutions are sorted according to participants' preferences to indicate who has priority to get the asset.
For example,
if Alice's escrowed asset $a_1$ is transferred to Bob in \emph{swap1}
but transferred to Carol in \emph{swap2},
then Alice, Bob, and Carol rank \emph{swap1} and \emph{swap2} by preference. 

\paragraph{Circuits use negations to indicate preferences.} Suppose \emph{swap1} is preferred than \emph{swap2}, and the circuit for those swaps are $C_1$ and $C_2$, respectively. Each circuit also indicates the recipient when they evaluate to \true. To implement this priority, the circuit on the escrowed asset $a_1$ would be $(\neg C_1 \wedge C_2) \vee C_1$, indicating that if $C_1$ evaluates to \true, \emph{swap1} will be completed and \emph{swap2} will only be completed if $C_1$ evaluates to \false.

 \subsection{Detailed Construction}
 
\subsubsection{Market Clearing Phase}
First, participants express their exchange requirements as before. Taking predicates $P_x$ from a party $x$, there will be an addition restriction $r_x$ due to the fact that  multiple arcs represent the same asset. $r_x$ is defined as: for arcs that represent the same asset, at most one of those arcs can be \true. This can be expressed in the same way as predicates we define in \secref{predicates}. Then, the new predicate for $x$ is $$P_x^{new}=P_x\wedge r_x.$$  For convenience, $P_x^{new}$ is called $P_x$ from now on. The set of new predicates are called $\cP$.

We find assignments that satisfy $\phi(\cP)$. The solutions are denoted by $S=\{s_1, \cdots, s_i,\cdots, s_k\}$. We sort the solutions by inclusion, and organize them into a DAG, and find hashlocks for each solution as \ProtocolA{}, except the redundancy provider is defined differently.

\begin{definition}
 Suppose $s_1$ and $s_2$ are two solutions for a predicated graph $(\cP,G)$.
  A party $x$ is a redundancy provider in \ProtocolB{} iff, it is a redundancy provider defined in Def. \ref{redundancy_provider} and completing both $s_1$ and $s_2$ does {\bf{not}} conflict with $r_x$.
\end{definition} 

The reason why the definition is updated is that, if $s_1$ and $s_2$ shares the same asset with different recipients(i.e. completing both $s_1$ and $s_2$ conflicts with $r_x$), it is impossible to complete both.  In other words, $x$ does not provide redundant collateral in $s_1$ and $s_2$. It is not a redundancy provider in this case.

In addition to sorting solutions into a DAG by inclusion,
we also sort them by participants' preferences.
Assume there is a protocol which allows the participants
to agree on a ranking: a total order on the solutions.
Let $S^*:=\{s^*_1,s^*_i,\cdots, s^*_k\}$ be the set of sorted solutions,
where $s^*_i$ \emph{precedes} $s^*_j$ if $i<j$.
In other words, $s^*_i$ is preferred over $s^*_j$.
To distinguish, we call the circuit $C_{s^*_i,q}$ in the original swap scheme as old circuit, the one in \ProtocolB{} as new circuit.
For each swap scheme $(s^*_j,q)$,
the new hashlock circuit is \begin{equation*}
C^{new}_{s^*_j,q} := (\bigwedge _{i< j, \forall q'\in Q(T(S)),s^*_j \text{ conflicts with } s^*_i }\neg C_{s^*_i,q'}) \wedge C_{s^*_j,q}.
\end{equation*}
The new circuit implements the following logic: a swap $s_j^*$ is completed if and only if the hashlocks in $s_j^*$ are unlocked, and there is no preceding conflicting swap $s_i^*$($i<j$) such that hashlocks of $s_i*$ that can be unlocked.

\subsubsection{Running the protocol on chain}
 
This phase is similar to the previous protocol,
but it only include two phases: \emph{Escrow Phase} and \emph{Redeem Phase}.

\paragraph{Escrow Phase}

Each participant runs $\Swap(G_{s_i}, \cH_{s_i,q}).\Escrow$. Note that the circuit corresponding to $\cH_{s_i,q}$ is $C_{s_i,q}^{new}$ now as defined above.
If an asset $a$ is already escrowed,
$C(a):= C_{current}(a) \vee C_{s_i,q}^{new}$. Suppose $C_{\current}(a)$ means the current hashlock circuit on an arc $a$.
If an asset participates in multiple swaps with different recipients,
the $\cH_{s_i,q}$ also specifies the recipient in this swap.

\paragraph{Redeem Phase}
 We say a hashlock set $\cH_{s_i}$ is unlocked when all hashlocks in $\cH_{s_i}$ are unlocked, and a hashlock set $\cH_{s_i}$ times out if any hashlock in $\cH_{s_i}$ times out. We cannot let parties simply run $\Swap(G_{s_i},  \cH_{s_i,q}).\Redeem$ since it is not safe. For example, if a party's one outgoing arc has  $\cH_{s_i}$ unlocked, and all $\cH_{s_j}$ where $j<i$ times out, then this outgoing arc will be triggered as in $\Swap(G_{s_i}, \cH_{s_i,q})$.
However, if another outgoing arc has $\cH_{s_j}$ unlocked where $j<i$, then this party's incoming arcs will have $\cH_{s_j}$ unlocked, then all incoming arcs can be triggered in swap scheme $\Swap(G_{s_j}, \cH_{s_j,q})$. Completing payments in two different schemes may produce a worse payoff. 
To overcome this problem,
we use to a broadcast scheme to synchronize the state of hashlocks.
Assume there is a broadcast scheme where there is an upper bound $t_u$ to
synchronize all hashlocks such that,
if a hashlock $h$ is unlocked on any arc,
then all arcs' hashlocks $h$ can be unlocked.
The redeem phase takes  $(MaxPathLength(G) \Delta +t_u)$, where $MaxPathLength(G)$ is the length of the longest path in the graph.
We provide a broadcast scheme based on a modification of hashkeys.

The key behind our design is, once a hashkey appears on any arc of a party,
this party can relay it to all its related arcs,
both outgoing arcs and incoming arcs.
We first transform the directed graph $G$ into an undirected graph $G^u$.
If there is more than one arc between two vertices,
we just add one to the undirected graph. 
Then, a hashkey corresponds to a simple path in the undirected graph.
It times out after $(MaxPathLength(G)+|p|)\Delta$,
where $G$ is the original directed graph containing all participants,
and $p$ is a path in the transformed undirected graph $G^u$.
The Redeem phase ends after $(MaxPathLength(G)+MaxPathLength(G^u))\Delta$.

An asset transfer happens on an arc if one clause of its hashlock circuit is \true{}. 
 Each clause corresponds to hashlock circuits in one swap scheme, which is composed of the negations of conflicting preceding solutions' (old) circuits and the current solution's (old)hashlock circuit.
The clause is \true{} only if both of the following two conditions are met.
\begin{itemize}
    \item All conflicting preceding solutions' old circuits are \false{} until timeout.
     That means the contract needed to wait until it timed out to decide whether conflicting preceding solutions' old circuits are unlocked.
     \item Current solution's old hashlock circuit need to evaluate to \true.
      That means all hashes in the old hashlock circuit need to be unlocked before they time out. 
\end{itemize}

All contracts agree on the order of conflicting solutions.
A solution is triggered only if the conflicting preceding solutions are not triggered.
Each arc will at most complete one asset transfer to one recipient.
A detailed analysis appears in \secref{properties}. %

\subsection{Ranking Solutions}
Solutions that can conflict with each must be ranked to decide which one is preferable.
This ranking is established by some kind of negotiation.
Participants who are not proposers can accept if the proposed order is acceptable and leave the deal otherwise. We can design a more sophisticated protocol to make a smarter decision. Details are left to interested readers.

\section{Analysis and Proof}
\seclabel{properties}
In this section, we analyze our proposed protocols. We provide definitions of required properties below, and  detailed proof for those properties can be found in \secref{proof}.

\subsection{Security Properties}
Let $S$ be the set of solutions output by an all-SAT solver.
Here are some desired properties for both protocols.

\begin{definition}
\emph{Universal Liveness}:
if all parties are conforming,
then a maximal set of compatible exchanges out of $S$ can be completed. 
\end{definition}

\begin{definition}
\emph{Local Liveness}:
in a swap scheme $(G,\cH)$,
if all involved parties are conforming,
and this scheme is selected (in \ProtocolA) or no scheme with higher priority is completed
(in \ProtocolB), then the asset transfers in $G$ can be completed.
\end{definition}

\begin{definition}
\emph{Safety}:
A conforming party $x$ will never end up worse off,
that is, $x$ never pays an asset unless its liveness predicate is \true, and never overpays (as defined by the safety predicate in \secref{predicates}). 
\end{definition}

\begin{definition}
\emph{Fault-tolerance}.
A party can complete an exchange according to its liveness predicate as long as there exists one swap scheme $s$ where all parties are conforming,
and $s$ is chosen by negotiation (in \ProtocolA)
or no scheme with higher priority is completed (in \ProtocolB).
If this party is involved in $m$ swap schemes
(out of the swap schemes output by an all-SAT solver),
we say they can tolerate $(m-1)$ failed schemes.
\end{definition}

\begin{definition}
A protocol is a
\emph{strong Nash equilibrium strategy}
if  no coalition improves its payoff when its members cooperatively deviate from the protocol.
\end{definition}

\begin{definition}
A protocol provides
\emph{economic-efficient redemption}
if a party can complete multiple exchanges by only redeeming once.
\end{definition}

\subsubsection{Customized Properties for \ProtocolA}

\begin{definition}
A protocol satisfies
\emph{fast settlement} if it has no hard timeout.
If all parties are compliant,
no party has to wait a fixed time to complete the transfer.
\end{definition}
\subsection{Protocol Comparisons }
We compare the two proposed protocols with prior protocols in \secref{comparison}.

\section{Related Work}
\seclabel{related}
There is an extensive body of research on blockchain interoperability \cite{belchior2021survey}.
Some research addresses general-purpose cross-chain communication,
focusing on the problem of reliably communicating the source chain's internal state
to a target chain.
Other research addresses atomic asset transfers,
for example, Alice trades her bitcoin for Bob's ether.

This paper's protocols focus on cross-chain asset transfers.
Note that we do not mint or burn any assets.
An asset transfer occurs between a sender and a receiver.
There are many prior proposals for asset exchanges.
Some are centralized~\cite{heilman2020arwen},
and some use connectors to route packetized payments across different blockchains~\cite{thomas2015protocol}.
These protocols are surveyed elsewhere~\cite{robinson2021survey,belchior2021survey}.

Many prior works utilize \emph{hashed timelock contracts} (HTLCs).
HTLCs allow one party to safely exchange assets with another party
without the need for a trusted third party.
The only trust anchors are the blockchains themselves,
and any blockchain that supports smart contracts supports HTLCs.
HTLCs are the basis for the Lightning network~\cite{poon2016bitcoin},
for two-party atomic cross-chain swaps~\cite{tiernolan,decred,sparkswap},
and for multi-party atomic cross-chain swaps~\cite{Herlihy2018} on strongly-connected digraphs.
In \cite{shadab2020cross},
the authors integrate off-chain steps to deal with swaps whose digraphs may
not be strongly connected.

Cross-chain transactions that tolerate deviating participants are studied
by Bagaria \emph{et al.}~\cite{bagaria2020boomerang},
which proposes a technique called \emph{Boomerang} to be used on top of multi-path routing schemes to construct redundant payment paths.
A payment is split into $n$ paths,
each of which carries the same amount of payment.
The payee can trigger $t$ paths out of $n$,
where $t$ yields the intended amount of payment.
If the payee tries to cheat by collecting payments from more than $t$ paths,
all payments will be voided.
The approach can tolerate participants on $(n-t)$ paths being faulty.
A limitation of this approach is that it has to split the payments to $n$ equal shares
and it does not support heterogeneous payments along different paths.
\emph{Spear}~\cite{rahimpour2021spear} improves \emph{Boomerang} by allowing different
payment amounts on different paths.
However, Spear only works for single payments from one payer to one payee,
e.g. from Alice to Bob, and the multi paths are disjoint.
It is not straightforward to generalize these protocols to a cyclic graph,
multiple paths overlap, and multiple parties have set up tentative redundant payments.
Mercan \emph{et al.}~\cite{mercan2021improving} propose protocols to improve
transaction success rate in payment networks by improving payment channel networks' performance
with better routing strategies and ways to address imbalances.

\section{Remarks}

Conclusions and discussions can be found in \secref{remarks}.

\bibliographystyle{ACM-Reference-Format}
\bibliography{zotero,references}  
\clearpage
\newpage

\appendix
\section{Appendix}
\subsection{Proof}
\seclabel{proof}

\subsubsection{Proof for the base protocol}

We first prove properties of the base protocol–the swap scheme. Our base protocol is a descendant of Herlihy's swap protocol \cite{Herlihy2018}, with one change: redundancy providers are asked to provide a hashlock on all arcs. Here we show that our base protocol inherits properties of the original protocol.

\begin{theorem}
\thmlabel{theo:base}
Our base protocol guarantees: (1) if all parties are conforming, then the swap in the graph happens. (2) a conforming party is always safe:  i.e. they never pay any assets without getting all entering assets in the graph.
\end{theorem}

\begin{proof}
If all parties are conforming, then the assets in the graph will be escrowed starting from leaders, since the graph left by removing leaders' arcs is a directed acyclic graph, and there is a order to traverse all vertices by topological sorting. When a vertex is traversed, it can escrow outgoing assets. 
After the escrow phase, no matter a party is a leader or a follower, once a hashkey unlocks a hashlock $h$ on its outgoing arc, it can generate a new hashkey and send it to its incoming arc to unlock the same hashlock. That means, if any hashlock on its outgoing arcs is unlocked,  then those  hashlocks  on its  incoming arcs will also be unlocked. In other words, if any outgoing arc is triggered, all of its incoming arcs can be triggered. Thus, a conforming party will never end worse off.

The redundancy providers are treated as additional leaders who send hashkeys corresponding to hashlocks generated by them after they receive all incoming escrows. The only difference between leaders and redundancy providers is that redundancy providers wait for all incoming escrows and before they their outgoing assets. 
\end{proof}

\subsubsection{Proof for \ProtocolA{}}

Here we prove properties of \ProtocolA{}.
\begin{theorem}
\emph{ProtocolA} satisfies liveness.
\end{theorem}
\begin{proof}
Suppose  there are $k$ solutions output by all-SAT solvers. If all parties are conforming, then the escrow phase in each solution is completed. Then,  in the $Select$ phase, a conforming redundancy provider proposes a maximum set of solutions $S_c$ where no solutions are mutually exclusive. Since those solutions are compatible with each other, other conforming redundancy providers vote yes to accept those solutions. Then for each solution $s\in S_c$, the redundancy providers and leaders release hashkeys in the $Redeem$ phase, and other parties redeem when they receive hashkeys on outgoing arcs. All assets will be redeemed by \thmref{theo:base}. The swaps in chosen solutions can be completed. When a swap happens, the involved parties' liveness predicates are satisfied.
\end{proof}

\begin{theorem}
\label{theorem:safe}
In \emph{ProtocolA}, a conforming party is always safe.
\end{theorem}
\begin{proof}
Recall that any swap scheme individually guarantees all involved parties are safe: all-or-nothing.  We just need to care about what happens if a party is involved in more than one swap schemes. Here we show that \emph{ProtocolA} keeps them safe with regard to their safety predicates.

For any solution $s_i$, no matter it is chosen or not, a party is playing a role in the base protocol.  Since we reuse some secrets in solutions $s_i\subsetneq s_j$, we cannot analyze the safety independently. The reuse of secret poses safety at risk since if a hashkey is released, it can also be used in another swap scheme, rather than the swap scheme that they intend to complete. Thus, we need to analyze the safety implications of reusing secret keys. Recall that when we reuse secrets, a solution $s_j$ uses secrets from $s_i$ if $s_i\subsetneq s_j$.

For any scheme $(G_{s_i},\cH_{s_i,q})$, no matter it is chosen or not, this party is safe by the following analysis.

\begin{enumerate}

        \item If this party $x$ is a redundancy provider or a leader,  then this party only releases the secret $h \in \cH_{s_i,q}$ in $(G_{s_i},\cH_{s_i,q})$ when he/she receives all incoming escrows in this solution. Because of the reuse of secrets, it is possible that incoming assets in $(G_{s_i},\cH_{s_i,q})$ are escrowed, however, incoming assets in other graph $(G_{s_j},\cH_{s_j,q})$, where $h\in \cH_{s_j,q}$, are not escrowed but outgoing assets in $(G_{s_j},\cH_{s_j,q})$ are escrowed. When he/she releases this secret, it is possible $\cH_{s_j,q}$ is unlocked and the outgoing arcs are triggered without all incoming arcs being triggered. In the \emph{redeem} phase, to avoid unsafe outcomes, a conforming redundancy provider or a leader will not release the secret for that case.  By not releasing the hashkey that is reused in multiple schemes when some of them have deviating parties, a party does not risk losing assets without getting expected assets. \label{case:leader}
    
  \item If this party $x$ is neither a redundancy provider nor a leader, this party does not own a secret and it just propagates secrets from leaders. Since this party only escrows after receiving incoming escrows, that guarantees him/her that,  if any outgoing arc is triggered using $\cH_{s_i,q}$ by a set of hashkeys, he/she can always construct new hashkeys on the base of the aforementioned hashkeys to trigger all incoming arcs. The outcomes for $x $ in a single scheme is all-or-nothing, where all means all incoming assets are triggered, and nothing means no outgoing assets are triggered.  In either case, $x$ is safe.
\end{enumerate}

In a nutshell, if a party is neither a redundancy provider nor a leader, then each exchange can be completed independently and treated separately. The safety property is preserved. If any of exchanges are completed, the liveness predicate is \true{}. If all exchanges fail, then this party gets all assets escrowed in all exchanges refunded.

If a party is a redundancy provider or a leader, it releases secrets only when releasing it has no risk of completing a partial swap.
\end{proof}

\begin{theorem}
If a party is involved in $m$ schemes in the market clearing phase, then with our protocol, he/she can tolerate failure of up to $m-1$ schemes.
\end{theorem}

\begin{proof}
If a party is involved in $m$ schemes, and there is a scheme where all parties are conforming in the escrow phase, then the protocol can proceed to the redeem phase. If the correct solution is chosen, then he/she can complete the exchange, making his/her liveness predicate \true{}.
\end{proof}

\begin{theorem}
\ProtocolA{} is a strong Nash equilibrium strategy.
\end{theorem}
\begin{proof}
We prove that no coalition has the incentive to deviate if they assume participants outside the coalition are conforming. We prove by contradiction. Suppose there is a coalition of participants that improve their payoff by deviating. This coalition must get more than what they should or pay less than or both. In any case, there must be a victim who pays more or gets less or both. In either case, a victim ends with a payoff that contradicts its safety predicate. Since a conforming party is always safe by Theorem \ref{theorem:safe}, they do not overpay or get less, which gives us a contradiction.
\end{proof}

\begin{theorem}
Participants can redeem their incoming assets in multiple schemes with a scheme corresponding to the smallest graph they participate in.
\end{theorem}

\begin{proof}

Due to reuse of hashes, participants can release one hashkey to unlock the same hash used in multiple schemes. If a party releases a hashkey corresponding to the smallest graph they are involved in, the hashkey can be applied to unlock hashlocks in larger graphs that include this small graph. Thus, participants only need to redeem their incoming assets with the scheme of the smallest graph.
\end{proof}

\begin{theorem}
Participants can settle a swap immediately after the escrow phase of a swap scheme is completed if all parties are conforming and respond quickly.
\end{theorem}

\begin{proof}
After the escrow phase of a swap scheme completes, leaders and redundancy providers can choose to release secrets to redeem the assets. As long as they did not agree to complete mutually exclusive schemes before, they can agree on this swap scheme and complete the redeem phase.
\end{proof}

\subsubsection{Proof for \ProtocolB{}}

The proof for common properties liveness, a strong Nash equilibrium strategy, redeem efficiency, fault-tolerance are similar to the proof we have for \ProtocolA{}. Here we focus on the difference: safety.

There is a unique total order of all possible swap schemes. Each swap scheme is assigned an index $i\in[1,m]$ where $m$ is the number of distinct swap schemes.  $s^*_i$ has higher priority than $s^*_j$ if $i<j$. We first define a state called \emph{level} to denote which asset transfer will be triggered on an arc, i.e. who is the recipient.
\begin{definition}

If an arc has its hashlock circuit evaluates to \true{}, where the clauses with highest priority that evaluate to \true{} corresponding to the swap scheme $s^*_{i}$, then we say the arc is at level $i$. 
\end{definition}

\begin{definition}

If an arc has its hashlock circuit evaluates to false, the arc is at state $Refunded$, we define it as level $\infty$ for consistency.
\end{definition}

Assuming escrow phase is executed without deviation, we prove two things:

\begin{theorem}\label{theorem:level i}
For a conforming party, if any of its outgoing assets is at a level $i\geq 1$, then all of his/her escrowed incoming assets and escrowed outgoing assets are at level $i$.
\end{theorem}

\begin{proof}

If an arc is at level $i$, that means the the clause corresponding to swap scheme $s^*_{i}$ evaluates to \true{}. The hashlocks in $s^*_{i}$ are unlocked, and hashlocks on all $s^*_{j}$ $j<i$ are not unlocked.  Based on the property of hashkey schemes which uses undirected paths, this party can unlock hashlocks corresponding to $s^*_{i}$ on all its incoming assets and outgoing assets in that scheme. 
\end{proof}

\begin{theorem}
A conforming party is safe.
\end{theorem}
\begin{proof}
Due to Theorem \ref{theorem:level i}, this party can complete a trade with level $i$. No outgoing assets will be at a different level. Each level depicts a safe swap scheme, thus this party is safe.
\end{proof}

\subsection{Comparison among \ProtocolA{} and \ProtocolB{} and Previous Protocols}
\seclabel{comparison}
\subsubsection{Comparison between \ProtocolA{} and \ProtocolB{}}
\begin{itemize}
    \item Time efficiency. \ProtocolA{} does not require a hard timeout while \ProtocolB{} does,
implying that \ProtocolA{} can be settled faster than \ProtocolB.
    
\item Collateral. \ProtocolA{} does not allow one escrow to have multiple possible recipients while \ProtocolB{} does,
  implying that \ProtocolB{} needs lower levels of collateral.
    
\item Fault-tolerance. \ProtocolB{} provides more fault-tolerance than \ProtocolA.
  Both tolerate deviating parties in the \emph{Escrow Phase}. As long as there is a swap whose escrow phase finishes, the redeem phase can be continued.
  However, only \ProtocolB{} tolerates deviating parties in the \emph{Redeem Phase}. In \ProtocolA{}, after \emph{Escrow Phase}, they need to select a subset of schemes to proceed. If the participants in the selected schemes abort in the \emph{Redeem Phase}, then the swap fails. On the contrast, in \ProtocolB{}, they do not need to reach an agreement to select a scheme after \emph{Escrow Phase}. They can proceed with all schemes whose \emph{Escrow Phase} finish, and redeem in all those swaps. \ProtocolB{} provides more swap schemes than \ProtocolA{} to complete. Thus, it is more fault-tolerant.

\item Flexibility.
  In \ProtocolA, participants can escrow different assets to trade,
  giving the participants more chance to find a counterparty to a trade.
  In the Redeem Phase,
  if a participant is a leader,
  they can also choose the most economically efficient trade to complete.
  Thus, \ProtocolA{} is more flexible.
\end{itemize}

\subsubsection{Comparison with Previously Protocols}
\seclabel{comparison_with_previous}

Here we provide a comparison in terms of time needed for our proposed protocols and previous protocols.

We do not count in the computation time of off-chain processes, such as finding feedback vertex sets, finding solutions, determining the set of redundancy providers and reaching agreement. Those processes are efficient since the problems are trivial and they are not bounded by the throughput of blockchain. 

We focus on the time consumption on-chain. Previously protocols do not support robustness and participants have to try different alternatives sequentially. Suppose each alternative swap takes $m\Delta$ to expire, $\Delta$ is the upper bound of message delay. When we say a swap expires, we mean the assets are refunded because of failure.\footnote{Different assets in the swap are refunded at different time. Here we use the time when last asset expires for brevity.} Let $\epsilon$ be the time needed in the \emph{Escrow Phase} and \emph{Redeem Phase} to complete a successful swap.  $\epsilon$ can be much small compared to $\Delta$ if participants execute the swap protocol quickly.

Suppose each swap has an independent probability of $q$ to fail. If a participant tries sequentially, then the expected number of tries is $\frac{1}{q}$. The time needed is $max\{(\frac{1}{q}-1)m\Delta,0\}+\epsilon$. For example, if $q=0.25$, the time needed is  $3m\Delta+\epsilon$.

Let $\gamma$ denote the time needed for finding solutions in \emph{market clearing phase}. If we try those alternative in parallel with \ProtocolA{}, in the best case, the time needed is $\epsilon+\gamma+\omega$, where $\omega$ is time needed for \emph{Select phase}. Note that the \emph{market clearing phase} for finding solutions and the \emph{select phase} to choose solutions can be short since they are not bounded by the throughput of blockchains, thus much less than $\Delta$. In the worst case, the time needed is $m\Delta+\gamma$ meaning the successful swap finishes right before it expires.

With \ProtocolB{}, if they do not complete the most prioritized swap,  in the both best and  worst case, the time needed is $m\Delta+\gamma+\omega'$ where $\omega'$ is time needed for \emph{sorting the solutions}.  If they complete the most prioritized swap, then the time needed is $\epsilon+\gamma+\omega'$ in the best case since there is no waiting needed for check the status of a conflicting but more prioritized swap.

\subsection{Remarks}
\seclabel{remarks}
This paper explores trade-offs among time, fault-tolerance, and collateral.
As illustrated by \ProtocolA,
multiple escrows may incur costs when blockchains and counterparties charge fees.
On the other hand, as illustrated by \ProtocolB,
eliminating duplicate escrows can introduce delays in the form of hard timeouts.

We use the notion of a predicate to capture the complexity of multi-party swaps.
Prior safety definitions~\cite{Herlihy2018} tend to be too restrictive,
requiring, for each party,
that all outgoing assets to be refunded if any incoming asset is not redeemed.
Our use of predicates captures the notion that certain subsets of the potential
trades are satisfactory.

The hashkey mechanism in the base swap scheme uses path signatures, which require signatures of parties included in the path. We extend the use of path signatures~\cite{Herlihy2018}
to allow the set of signatures from parties that do not necessarily for a path, using \emph{pathless} hashkeys. 
Here, a pathless hashkey can correspond to any set of parties' signatures as long as there is no duplicated parties in the signature. This change can provide more fault-tolerance since if as long as leaders release their secrets,
other parties, say $u$ can redeem their incoming assets. In this way, $u$ does not depend on parties on the path from $u$ to the leader to sign and relay the hashkey. More details are provided in \secref{shortcut_hashkey}.

\subsubsection{Discussion on Maximum Set of Schemes}
\seclabel{maximum}
A set of schemes is maximal, if no other schemes can be added to the set without conflicting with existing schemes. This definition does not guarantee to satisfy as many parties’ liveness predicate as possible. For example, Alice has two conflicting alternative swaps: one is to exchange assets with Bob(2 party swap), and the other is to exchange assets with David through Carol as an intermediary(3 party swap). Even if all of them are conforming in the escrow phase of each swap scheme, at last only one swap will be completed. Either swap is considered the maximal, even though only 2 parties are satisfied in 2 party swap, less than 3 party swap. The reason why we define a maximal set of schemes in this way is that we focus on completing a swap scheme as long as it does not conflict with what they have agreed on. The select phase can be executed gradually, and a group of participants in a swap scheme can complete the swap as soon as their escrow phase finishes if the swap scheme does not conflict with any previously completed swaps.  

It is interesting  to have more sophisticated mechanisms to try to satisfy as many participants as possible. We will leave this to future work.

\subsubsection{Consideration on Number of Alternatives }
\seclabel{number_k}
The number of  alternative solutions can be decided by participants based on their fault tolerance preferences and cost of running those alternatives , e.g. if a participant wants more fault-tolerance, he/she may want to have more alternatives, at the cost of paying more transaction fee or commission to parties who wants to trade with him/her. 

\subsubsection{Complex Predicates}
\seclabel{appendix_pred}
For complex predicates such as ``At least $k$ out of $n$ incoming arcs'' or ``At most $k$ out of $n$ outgoing arcs'', 
it might seem that expressing these conditions one would need exponentially (e.g. $n \choose k$) clauses in conjunctive normal form (CNF).
However, it is actually not the case. 
Those predicates can be efficiently expressed using boolean circuits, which can be converted to CNF in linear size
of the circuit (also known as Tseytin transformation~\cite{tseitin1983complexity}). The resulting CNF can be fed to SAT solvers.

\subsubsection{More Fault-tolerance with Pathless Hashkeys}
\seclabel{shortcut_hashkey}

A \emph{pathless} hashkey for arc $(u,v)$ is a variation where, given $(s,p,\sigma)$,
$p$ can be any set of unique vertices of $G$ (not just a path).

Here we compare the original hashkey mechanism \cite{Herlihy2018} with our proposed pathless hashkey mechanism and answer some questions that may be interesting to readers.

Hashkeys associated with paths and signatures are proposed in Herlihy's swap \cite{Herlihy2018}. In this paper, we propose pathless hashkeys which are similar to hashkeys and the only difference is that a pathless hashkey does not have to be associated with a path. It can be associated with a set of vertices as long as those vertices are in the graph and there is no duplicated vertices. The main consequence of using pathless hashkey is that the duration of the redeem phase is prolonged. With traditional hashkeys, the assets are locked until $2MaxPathLength(G)\Delta$ after the start of the protocol execution. With pathless hashkeys, the assets are locked until $(MaxPathLength(G)+n)\Delta$, see reasons below.

\paragraph{Time Complexity}

In the original swap scheme, the assets are locked until $2MaxPathLength(G) \cdot \Delta$, since escrow and redeem both takes $MaxPathLength(G)$ rounds. If we use pathless keys, the escrow phase still takes $MaxPathLength(G)$ number of rounds since the graph does not change, but the redeem phase will take $n \Delta$, since every deviating coalition (say $n-1$ parties) can send a hashkey in $(MaxPathLength(G)+(n-1))$ rounds, and the conforming party needs one more round to sign on it, and propagate the hashkey to its outgoing arcs. That means $(MaxPathLength(G)+n)$ rounds are required for safety purposes. In total, assets are locked until $(MaxPathLength(G)+n)\Delta$.

\subsubsection{Discussion on Shortcuts}

The hashkey mechanism requires parties in the path to be conforming to provide their signatures. If a party on the path is deviating, some parties’ asset transfer  cannot be completed because of the missing of this party’s  signature. Beside pathless hashkeys, another possible solution to tolerate deviating parties in the path is to use a shortcut of hashkeys. If the signature of a party on the path, the shortcut hashkey can be used to unlock a hashlock. We show why this approach does not work in this section.

Suppose a hashkey is associated with a path $p=(u_0,\cdots,u_k)$. A shorcut hashkey is defined as a hashkey whose path $p'=(u_0',\cdots,u_m')$ satisfies (1) $|p'|\leq |p|$ (2) all nodes in $p'$ are also in $p$, (3) and the order of nodes in $p'$ is consistent with their order in $p$.
A shortcut of a hashkey allows parties to redeem using less intermediate nodes' signatures on the path. In other words, a hashkey does not have to corresponds to a complete path. It can skip some nodes that should be in a complete path. For example, in a graph showing in \figref{fig:swap_graph}, $A$ is a leader. All valid paths regarding hashkeys on each arc is shown on the arcs, as $p$. If we allow shortcut of hashkeys to unlock a hashlock, then, there is an attack shown in \figref{fig:swap_graph_problem}. In \figref{fig:swap_graph_problem}, the blue paths $p$ denote the path of hashkeys to unlock a hashlock in the original swap scheme. We ignore the signatures and secrets in the hashkey for brevity and assume they are well-formed. In the original swap protocol, $A$ is a leader, and it first sends a hashkey with path $p=A$ to the arc $(B,A)$ to redeem. Then, after seeing this hashkey, $B$ appends itsself to the path $p$ (insert $B$ in front of $p$) and sends a hashkey with path $p=BA$ to the arc $(C,B)$ to redeem \footnote{We write the path in a more compact way, without parenthesis for brevity.}. $C$ and $D$ follows this pattern and send hashkeys with $p=CBA$ and with $p=DCBA$ to their incoming arcs one after the other.

Let $p'$ denote the paths that would be composed by parties if we use hashkeys shortcuts. $p'$ is constructed in the same way as $p$. 

 If $A$ does not release hashkeys corresponding to $p'=A$ to redeem $(B,A)$ and instead he/she colludes with $C$, then an execution of the protocol would propagate hashkeys as shown in the \figref{fig:swap_graph_problem} starting from $C$ redeeming $(D,C)$ with $p'=CA$. And then $D$ will redeem the arc $(B,D)$ with $p'=DCA$ and $B$ then has to redeem $(C,B)$ with $p'=BDCA$ which cannot unlock the hashlock generated by $A$ since the longest valid path(in the original swap scheme) corresponding to that arc has $|p|=2$, which is less than $|p'|=4$. That means, if we allow a shortcut of a path, then there is a problem since not all paths $p'$ are shortcuts of a original valid path $p$.

\begin{figure}
    \begin{subfigure}{.4\textwidth}
    \centering
     \includegraphics[width=\textwidth]{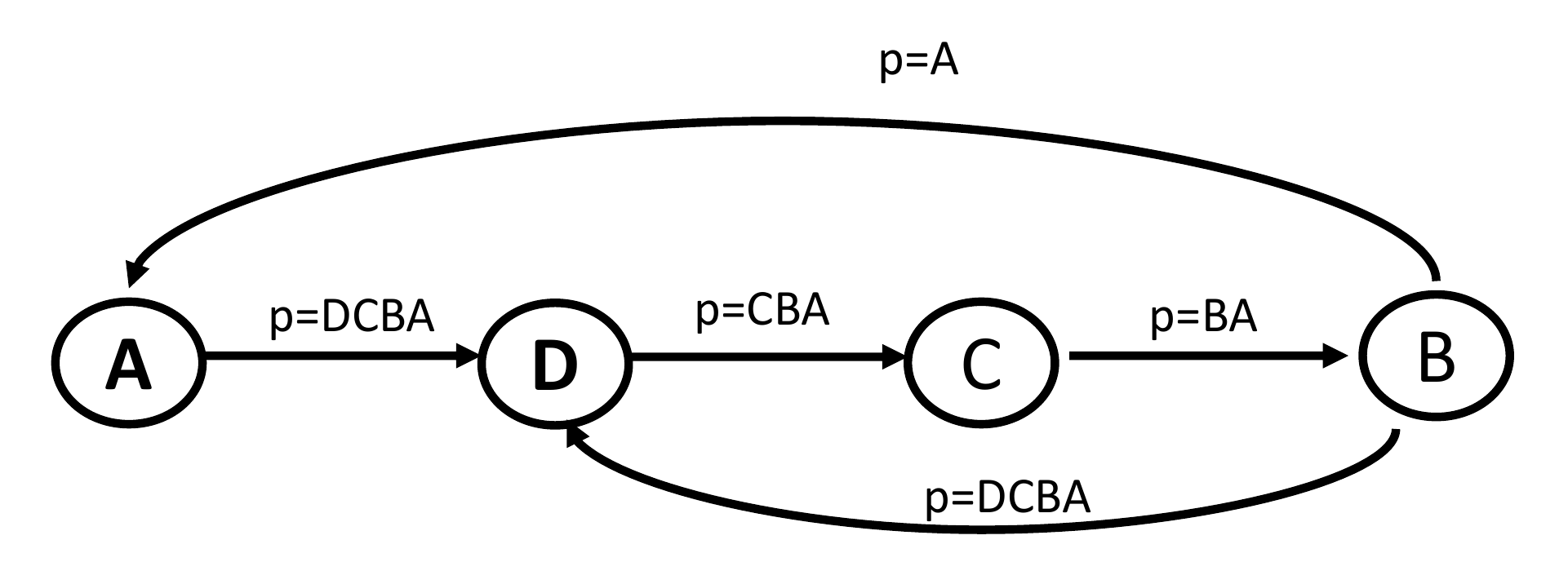}
    \caption{Paths of Hashkeys in a Swap Graph}
    \figlabel{fig:swap_graph}
    \end{subfigure}
   \begin{subfigure}{.5\textwidth}
   \centering
     \includegraphics[width=\textwidth]{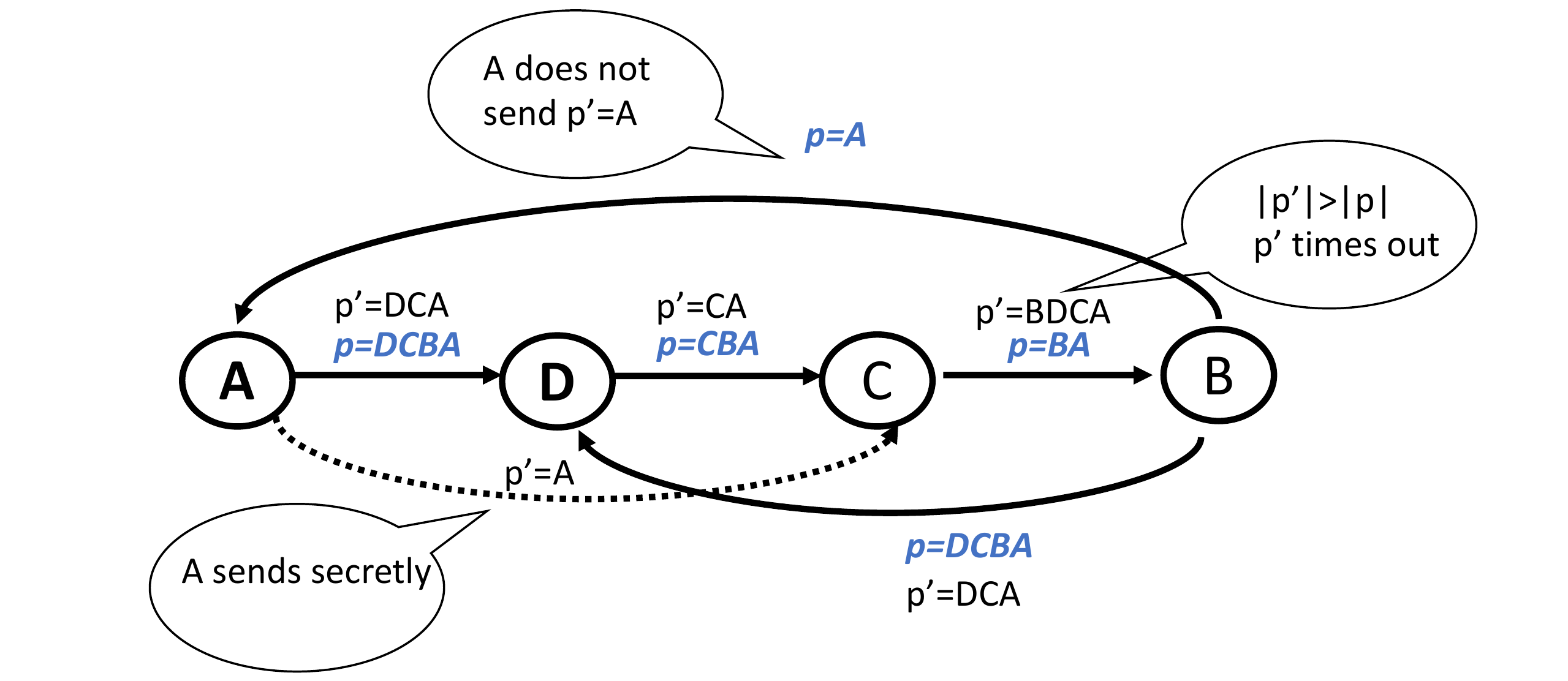}
    \caption{Shortcut of Hashkeys Does Not Work If $A$ Deviates}
     \figlabel{fig:swap_graph_problem}
     \end{subfigure}
     
         \caption{An Example Showing Why Shortcut of Hashkeys Does Not Work, Where $A$ Is a Leader}
\end{figure}

\begin{figure}
    \centering
     \includegraphics[width=.4\textwidth]{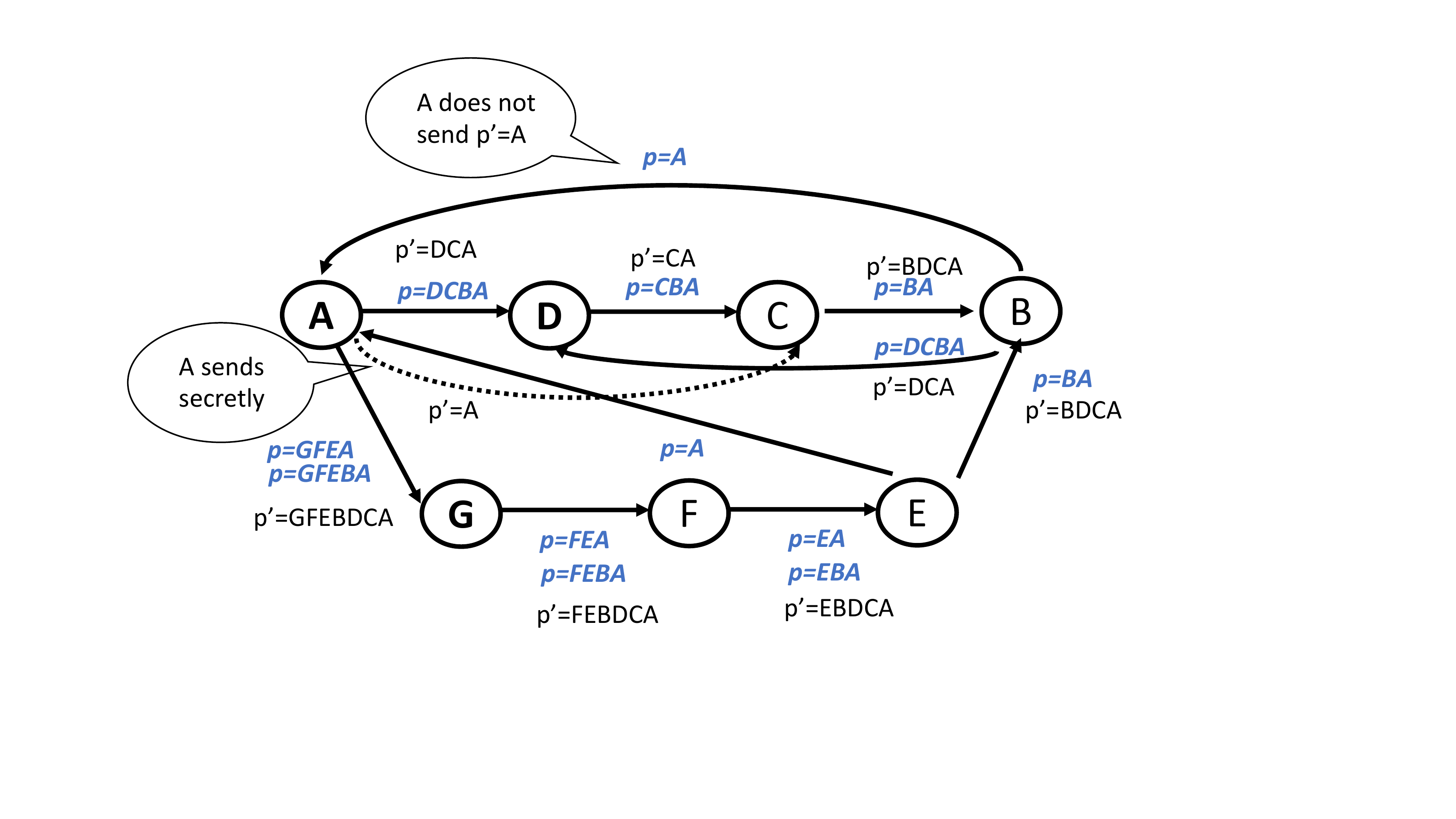}
    \caption{An example showing the timeout can be $n\Delta$ if all possible composition of hops are allowed}
    \figlabel{fig:timeout_long}
\end{figure}

Furthermore, we show that, if we fix the above problem by allowing a path longer than its original path to unlock a hash, then it is possible that we need to set the timeout to be $n \Delta$ for be safe, where $n$ is number of participants. As shown in \figref{fig:timeout_long}, if we do not allow shortcut of hashkeys, the original valid paths are shown by $p$. The new path $p'$ denotes paths composed by participants when redeeming. Suppose $A$ colludes with $C$ and send $p'=A$ secretly to $C$ and $C$ starts redeeming on $(D,C)$. If we allow a hashkey signed any combination of nodes as long as they do not form a circle, then the longest possible path's length from $A $ is $n=7$ on $(A,G)$. In the original scheme, the longest possible path is $5$ on $(A,G)$.

\end{document}